\documentclass[twocolappendix]{emulateapj}
\usepackage{amsmath}
\usepackage{float}
\usepackage{nicefrac}
\usepackage{rotating}
\usepackage{rotfloat}
\usepackage[none]{hyphenat}
\usepackage{enumitem}
\usepackage{afterpage}
\def\Lsun{L$_{\odot}$}
\def\Rsun{R$_{\odot}$}
\def\Msun{M$_{\odot}$}
\def\Zsun{Z$_{\odot}$}

\begin{document}

\title{A New Class of Nascent Eclipsing Binaries with Extreme Mass Ratios}

\author{Maxwell Moe\altaffilmark{1} \& Rosanne Di Stefano\altaffilmark{1}}

\altaffiltext{1}{Harvard-Smithsonian Center for Astrophysics, 60 Garden Street, MS-10, Cambridge, MA, 02138, USA; e-mail: mmoe@cfa.harvard.edu}

\begin{abstract}

Early B-type main-sequence (MS) stars ($M_1$ $\approx$ 5\,-\,16\,\Msun) with closely orbiting low-mass stellar companions ($q$ $=$ $M_2$/$M_1$ $<$ 0.25) can evolve to produce Type~Ia supernovae, low-mass X-ray binaries, and millisecond pulsars.   However, the formation mechanism and intrinsic frequency of such close extreme mass-ratio binaries have been debated, especially considering none have hitherto been detected.   Utilizing observations of the Large Magellanic Cloud galaxy conducted by the Optical Gravitational Lensing Experiment, we have discovered a new class of eclipsing binaries in which a luminous B-type MS star irradiates a closely orbiting low-mass pre-MS companion that has not yet fully formed. The primordial pre-MS companions have large radii and discernibly reflect much of the light they intercept from the B-type MS primaries ($\Delta I_{\rm refl}$ $\approx$ 0.02\,-\,0.14~mag).  For the 18 definitive MS\,+\,pre-MS eclipsing binaries in our sample with good model fits to the observed light curves, we measure short orbital periods $P$~=~3.0\,-\,8.5~days, young ages $\tau$ $\approx$ 0.6\,-\,8\,Myr, and small secondary masses $M_2$~$\approx$~0.8\,-\,2.4\,\Msun\ ($q$ $\approx$ 0.07\,-\,0.36).    The majority of these nascent eclipsing binaries are still associated with stellar nurseries, e.g. the system with the deepest eclipse $\Delta I_1$ = 2.8 mag and youngest age $\tau$~=~0.6\,$\pm$\,0.4~Myr is embedded in the bright H\,\textsc{ii} region 30 Doradus. After correcting for selection effects, we find that (2.0\,$\pm$\,0.6)\% of B-type MS stars have companions with short orbital periods $P$~=~3.0\,-\,8.5 days and extreme mass ratios $q$~$\approx$~0.06\,-\,0.25.  This is $\approx$10 times greater than that observed for solar-type MS primaries. We discuss how these new eclipsing binaries provide invaluable insights, diagnostics, and challenges for the formation and evolution of stars, binaries, and H\,\textsc{ii} regions.

\end{abstract}

\keywords{binaries: eclipsing, close; stars: massive, pre-main sequence, formation, evolution, statistics; ISM: H\,\textsc{ii} regions, evolution; galaxies: Large Magellanic Cloud}

\section{Introduction}

Close binaries with orbital periods $P$ $\lesssim$ 10$^3$ days are ubiquitous \citep{Abt1983,Duquennoy1991,Kobulnicky2007,Raghavan2010,Sana2012,Duchene2013} and are the progenitors of a variety of astrophysical phenomena \citep{Paczynski1971,Iben1987,Verbunt1993,Phinney1994,Taam2000}. Nonetheless, a close stellar companion cannot easily form in situ (see  \citealt{Mathieu1994} and \citealt{Tohline2002} for observational and theoretical reviews, respectively).  Instead, the companion most likely fragments from the natal gas cloud or circumstellar disk at initially wider orbital separations \citep{Bate1997,Kratter2006}.  Various migration hypotheses have been proposed for how the orbit decays to shorter periods \citep{Bate2002,Bonnell2005}.  These formation scenarios produce mostly close binaries with components of comparable mass because a low-mass companion either accretes additional mass from the disk, merges with the primary, remains at wide separations, or is dynamically ejected from the system.  

Close binaries with extreme mass ratios most likely require an alternative formation mechanism.  For example, a low-mass companion can be tidally captured into a closer orbit \citep{Press1977,Bally2005,Moeckel2007}, possibly with the assistance of gravitational perturbations from a third star~\citep{Kiseleva1998,Naoz2014}. Indeed, a significant fraction of close binaries are orbited by an outer tertiary \citep{Tokovinin2006}, suggesting the third star may play a role in the dynamical formation of the system.  It is fair to say that the mutual formation and  coevolution between massive stars and close companions are not yet fully understood. It has even been proposed that massive stars formed primarily via mergers of close binaries instead of through gas accretion from the circumstellar disk \citep{Bonnell2005,Bally2005}.  A complete census of close companions to massive stars must be conducted in order to determine the dominant formation mechanism of close binaries and massive stars as well as to reliably predict the production rates of certain channels of binary evolution.  

It is extremely difficult, however, to detect faint low-mass companions that are closely orbiting massive luminous primaries.  B-type main-sequence (MS) stars with low-mass secondaries have been photometrically resolved at extremely wide orbital separations $a$~$\gtrsim$~50~AU, i.e. long orbital periods $P$~$\gtrsim$~10$^5$~days \citep{Abt1990,Shatsky2002}.  Some of these resolved low-mass companions are still pre-MS stars that can emit X-rays \citep{Hubrig2001,Stelzer2003}.  Late B-type MS stars detected at X-ray wavelengths most likely have unresolved low-mass pre-MS companions at $a$ $\lesssim$ 50 AU \citep{Evans2011}.  However, the precise orbital periods of these putative X-ray emitting companions have not yet been determined. These unresolved binaries  may have short orbital periods $P$~$<$~10$^3$~days and may eventually experience substantial mass transfer and/or common envelope evolution as the primary evolves off the MS.  Alternatively, the binaries could have intermediate orbital periods $P$~=~10$^3$\,-\,10$^5$~days and could therefore avoid Roche-lobe overflow.

Multi-epoch radial velocity observations of double-lined spectroscopic binaries (SB2s) can provide the orbital periods $P$ and velocity semi-amplitudes $K_1$ and $K_2$.  Hence, the mass ratio $q$~$\equiv$~$M_2$/$M_1$~=~$K_1$/$K_2$ of an SB2 can be directly measured dynamically.  However, SB2s with MS components can only reveal companions that are comparable in luminosity, and therefore mass,  to the primary star.  SB2s with early-type primaries, known orbital periods, and dynamically measured masses all have moderate mass ratios $q$~$>$~0.25  \citep{Wolff1978,Levato1987,Abt1990,Sana2012}.   \citet{Gullikson2013} combined multiple high-resolution spectra of early-type stars in order to substantially increase the signal-to-noise. By implementing this novel technique, they detected SB2s with larger luminosity contrasts and therefore smaller mass ratios $q$~$\approx$~0.1\,-\,0.2. Although \citet{Gullikson2013} found a few candidates, stacking multiple spectra from random epochs in order to increase the signal-to-noise does not relay the orbital period of the binary. Similar to the case above of late-B primaries with unresolved, X-ray emitting companions, these SB2s with indeterminable periods may have wide orbital separations.  

Close faint companions to B-type MS primaries can induce small radial velocity variations, and these reflex motions have been observed with multi-epoch spectroscopy \citep{Wolff1978,Levato1987,Abt1990}. Although the orbital periods of these single-lined spectroscopic binaries (SB1s) can be measured, they have only lower limits for their mass ratios because the inclinations are not known. Nonetheless, an average inclination or a distribution of inclinations can be assumed for a population of SB1s in order to recover a statistical mass-ratio distribution \citep{Mazeh1992a}.  For SB1s with solar-type MS primaries $M_1$~$\approx$~1\,\Msun, the companions are almost certainly low-mass M-dwarfs \citep{Duquennoy1991,Mazeh1992b,Grether2006,Raghavan2010}.  For early-type MS primaries $M_1$ $\approx$ 10\,\Msun, however, SB1s can either contain $M_2$~$\approx$~0.5\,-\,3\,\Msun\ K-A type stellar companions or $M_2$~$\approx$~0.5\,-\,3\,\Msun\ stellar remnants such as white dwarfs, neutron stars, or black holes \citep{Wolff1978,Garmany1980}.  \citet{Wolff1978} even suggests that most SB1s with late-B MS primaries contain white dwarf companions, and therefore the fraction of unevolved low-mass stellar companions to B-type MS stars is rather small.  Unfortunately, there is at present no easy and systematic method for distinguishing between these two possibilities for all SB1s in a statistical sample.  Because early-type SB1s may be contaminated by evolved stellar remnants, it is prudent to only consider binaries where the nature of the secondaries are reliably known.  In addition to discovering close unevolved low-mass companions to B-type MS stars, we must also utilize a different observational technique for easily identifying such systems from current and future telescopic surveys. 

Fortunately, extensive visual monitoring of one of our satellite galaxies, the  Large Magellanic Cloud (LMC), conducted by the third phase of the Optical Gravitational Lensing Experiment (OGLE-III) has yielded a vast database primed for the identification and analysis of such binaries \citep{Udalski2008,Graczyk2011}.  OGLE-III surveyed 35 million stars in the LMC over seven years, typically obtaining  $\approx$470 near-infrared $I$ and $\approx$45 visual $V$ photometric measurements per star \citep{Udalski2008}.  Moreover, \citet{Graczyk2011}  utilized a semi-automated routine to identify more than 26,000 eclipsing binaries in the OGLE-III LMC database.  They cataloged basic observed parameters of the eclipsing binaries such as orbital periods $P$ and primary eclipse depths $\Delta I_1$, but the intrinsic physical properties of the eclipsing binaries still need to be quantified. 

  We previously showed that B-type MS stars with low-mass zero-age MS companions $q$~$\approx$~0.1\,-\,0.2 can produce shallow eclipses $\Delta I_1$~$\approx$~0.1\,-\,0.2~mag if the inclinations are sufficiently close to edge-on \citep[see Fig.~5 in][]{Moe2013}.  Indeed, the OGLE-III LMC survey is sensitive to such shallow eclipses, so we expect B-type MS stars with low-mass companions to be hiding in the OGLE-III LMC eclipsing binary catalog.   We therefore began to systematically measure the physical properties of the eclipsing binaries in hopes of identifying such extreme mass-ratio binaries.

  While  investigating the light curves of eclipsing binaries in the OGLE-III LMC database, we serendipitously discovered an unusual subset that displayed sinusoidal profiles between narrow eclipses (prototype shown in Fig.~1).  We soon realized the sinusoidal variations are caused by the reflection of light received by a large, low-mass, {\it pre-MS} companion from the hot B-type MS primary.  The present study is dedicated to a full multi-stage analysis of this new class of eclipsing binaries.   In \S2, we present our selection criteria for identifying ``reflecting'' eclipsing binaries with B-type MS stars and low-mass pre-MS companions.  We then measure the physical properties of these systems by fitting eclipsing binary models to the observed light curves (\S3).  We also examine observed correlations among various properties of our nascent eclipsing binaries, including their associations with star-forming H\,\textsc{ii} regions (\S4). In \S5, we correct for selection effects in order to determine the intrinsic frequency of close, low-mass companions to B-type MS stars.  In \S6, we discuss the implications of these eclipsing binaries in the context of binary star formation and evolution.  Finally, we summarize our main results and conclusions (\S7).

\begin{figure}[t!]
\centerline{
\includegraphics[trim = 1.5cm 0.1cm 0.2cm 0.5cm, clip=true, width=3.5in]{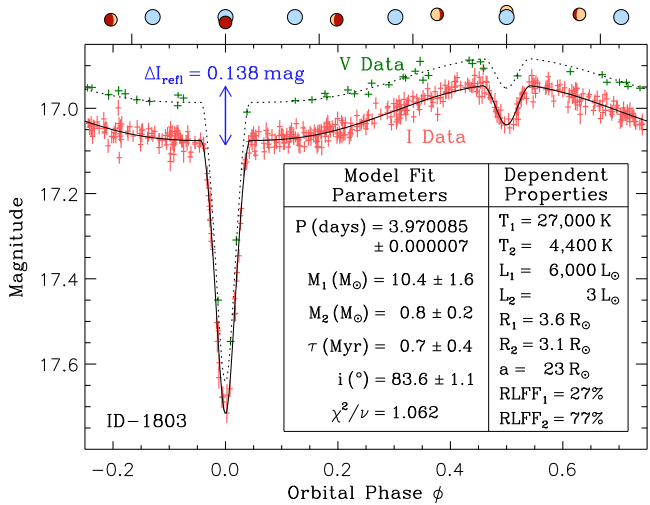}}
\caption{One of the 22 OGLE-III LMC eclipsing binary light curves with a B-type MS primary and low-mass pre-MS companion.  We fit a detailed {\it physical} model (black; see \S3 for details) to the V-band data (green) and I-band data (red).  The large reflection effect amplitude $\Delta I_{\rm refl}$ (blue) is used to identify such systems (see \S2).  Above is a to-scale schematic diagram of the binary at the orbital phases indicated by the tick marks.  The narrow eclipses dictate a detached binary configuration with Roche-lobe fill-factors $RLFF$ $<$ 80\%, which indicate both components are effectively evolving along their respective single-star sequences.  The inset tables show the main parameters constrained by the physical model fit (left), and the dependent properties (right) derived by using the model parameters in combination with Kepler's laws and stellar evolutionary tracks. The reported uncertainties in the physical model parameters are dominated by the systematic uncertainties in the stellar evolutionary tracks.  Because the distances to eclipsing binaries in the LMC are known, the photospheric properties are reliably measured from the observed light curve features, e.g.  $\approx$8\% uncertainties in the temperatures, $\approx$40\% uncertainties in the luminosities, and $\approx$10\% uncertainties in the radii, separation, and Roche-lobe fill-factors.  Note the extreme mass ratio $q$~=~$M_2$/$M_1$~=~0.07, young age, and how the primary B-type MS star is significantly hotter and more luminous than the pre-MS secondary.}
\end{figure}

\section{A New Class of Eclipsing Binaries}

\subsection{Selection Criteria and Analytic Models}

The OGLE-III LMC photometric database \citep{Udalski2008} lists the mean magnitudes $\langle I \rangle$, colors $\langle V-I \rangle$, and positions for each of the 35 million stars in their survey.    Throughout this work, we adopt a distance $d$ = 50 kpc to the LMC \citep{Pietrzynski2013}.  We also incorporate stellar parameters such as temperature-dependent bolometric corrections BC($T_{\rm eff}$) and intrinsic color indices $(V-I)_{\rm o}$\,($T_{\rm eff}$) from \citet{Pecaut2013}.  Based on these parameters, we select the ${\cal N}_{\rm B}$ $\approx$ 174,000 systems from the OGLE-III LMC catalog with mean magnitudes 16.0~$<$~$\langle I \rangle$~$<$~18.0 and colors $-0.25$~$<$~$\langle V-I \rangle$~$<$~0.20 that correspond to luminosities and surface temperatures, respectively, of B-type MS stars.

 The OGLE-III LMC eclipsing binary catalog \citep{Graczyk2011} provides the time $t$, photometric magnitude $I$ or $V$, and photometric error $\sigma_{\rm phot}$ for the ${\cal N}_I$ $\approx$ 470 and ${\cal N}_V$ $\approx$ 45 measurements of each eclipsing binary.  It also gives general properties of each eclipsing binary such as the orbital period $P$ (in days) and epoch of primary eclipse minimum $t_{\rm o}$ (Julian Date $-$ 2450000).  The orbital phase simply derives from folding the time of each measurement with the orbital period:

\begin{equation}
  \phi (t) = \frac{(t - t_{\rm o})\,{\rm mod}\,P}{P}.
\end{equation}

\noindent  We analyze the 2,206 OGLE-III LMC eclipsing binaries that have orbital periods $P$ = 3\,-\,15 days and satisfy our magnitude and color criteria. Such an immense sample of close companions to B-type MS stars is two orders of magnitude larger than previous spectroscopic binary surveys \citep{Wolff1978,Levato1987,Abt1990}.

 To automatically and robustly identify ``reflecting'' eclipsing binaries, we fit an analytic model of Gaussians and sinusoids to the $I$-band light curves for each of the 2,206 eclipsing binaries in our full sample. The parameters are as follows.  The average magnitude $\langle I \rangle$ is the total $I$-band magnitude of both stars if they did not exhibit eclipses or reflection effects.   The primary and secondary eclipse depths are $\Delta I_1$ and $\Delta I_2$, and the primary and secondary eclipse widths  are $\Theta_1$ and $\Theta_2$, respectively.  The phase of the secondary eclipse $\Phi_2$ provides a lower limit to the eccentricity of the orbit \citep{Kallrath2009}:

\begin{equation}
 e \ge e_{\rm min} = | e\,{\rm cos}(\omega) | = \pi | \Phi_2-\nicefrac{1}{2} |/2,
\end{equation}

\noindent where $\omega$ is the argument of periastron.  Finally, $\Delta I_{\rm refl}$ is the full amplitude of the reflection effect, which unlike eclipses, leads to an increase in brightness.  With these definitions, we model the I-band light curves in terms of Gaussians and sinusoids:

\begin{align}
I_{\rm GS}(\phi) = \langle I \rangle &+\Delta I_1 \Big[{\rm exp}\Big(\frac{-\phi^2}{2\Theta_1^2}\Big)+
           \rm{exp}\Big(\frac{-(\phi-1)^2}{2\Theta_1^2}\Big)\Big]\nonumber \\
  &+ \Delta I_2\, \rm{exp} \Big(\frac{-(\phi-\Phi_2)^2}{2\Theta_2^2}\Big) \nonumber \\
  &- \frac{\Delta I_{\rm refl}}{2}\Big[{\rm cos}\big(2\pi[\phi-\nicefrac{1}{2}]\big)+1\Big].
\end{align}

The photometric errors $\sigma_{\rm phot}$ provided in the catalog systematically underestimate the true rms dispersion outside of eclipse by (5\,-\,20)\% (see Fig.~2).  This is especially true for the brightest systems $\langle I \rangle$ $\approx$ 16.0\,-\,16.5 where the photometric errors $\sigma_{\rm phot}$ $\approx$ 0.008 mag are small.  We separately fit 3$^{\rm rd}$ degree polynomials across the out-of-eclipse intervals $3\Theta_1$ $<$ $\phi$ $<$ $\Phi_2 - 3\Theta_2$ and $\Phi_2 + 3\Theta_2$ $<$ $\phi$ $<$ $1-3\Theta_1$ for the eclipsing binaries with at least 50 data points across these intervals.  We then measure the rms dispersion $\sigma_{\rm rms}$ of the residuals resulting from these fits.   To rectify the differences between the catalog and actual errors, we multiply each of the photometric uncertainties by a correction factor $f_{\sigma}$:

\begin{equation}
 \sigma_{\rm corr}(t) = \sigma_{\rm phot}(t) f_{\sigma},
\end{equation}

\noindent  where $f_{\sigma}$ increases toward brighter systems:

\begin{equation}
  f_{\sigma}(I) = 1.05+0.15\times10^{(16.0-I)/2}.
\end{equation}

\noindent The source of this systematic error could partially be due to intrinsic variations in the luminosities of B-type MS stars at the 0.5\% level.  In any case, the corrected total errors $\sigma_{\rm corr}$ follow the measured rms errors $\sigma_{\rm rms}$ quite well (Fig.~2).  We implement these corrected errors $\sigma_{\rm corr}$ in our analytic light curve models below.

Even after accounting for the systematic error correction factor $f_{\sigma}$, a few of the photometric measurements are clear outliers.  We therefore clip up to ${\cal N_{\rm c}}$ $\le$ 2 measurements per light curve that deviate more than  4$\sigma$ from our best-fit model.  To be conservative, we only eliminate up to two data points to ensure we did not remove any intrinsic signals. Our analytic model has nine parameters (seven explicitly written in Eq. 3 as well as our own fitted values of $P$ and $t_{\rm o}$ according to Eq. 1), which provide $\nu$ = ${\cal N}_I - {\cal N}_{\rm c} - 9$ degrees of freedom. 

\begin{figure}[t!]
\centerline{
\includegraphics[trim = 1.7cm 0.7cm 0.0cm 1.1cm,clip=true,width=2.95in]{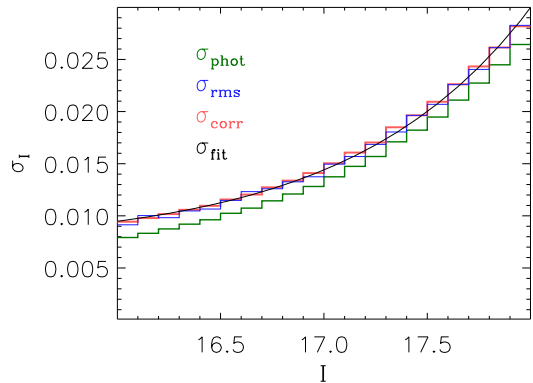}}
\caption{For each interval of I-band magnitudes, we compare the median values of the photometric errors $\sigma_{\rm phot}$ reported in the OGLE-III LMC eclipsing binary catalog (green), the intrinsic rms variations $\sigma_{\rm rms}$ outside of eclipse (blue), and the total corrected errors $\sigma_{\rm corr}$ (red).  We also display a fit $\sigma_{\rm fit}$ (black) to the total errors.}
\end{figure}

For the 2,206 eclipsing binaries, we use an automated Levenberg-Marquardt technique to minimize the $\chi_{\rm GS}^2$ statistic between the light curves and analytic models.  We also calculate the covariance matrix and standard 1$\sigma$ statistical uncertainties for our nine fitted parameters.  We visually inspect the solutions for all systems with $\chi_{\rm GS}^2 / \nu$ $>$ 1.5 to ensure the parameters converged to the best possible values.  For the few models that automatically converged to a local non-global minimum, we adjusted the initial fit parameters and reiterated the Levenberg-Marquardt technique to determine the lowest $\chi_{\rm GS}^2 / \nu$ value possible.

Our analytic model of Gaussians and sinusoids does not adequately describe some of the eclipsing binaries, which can lead to large values of $\chi_{\rm GS}^2/\nu$~=~2\,-\,5.  For example, some of our systems with nearly edge-on orientations exhibit flat-bottomed eclipses, and therefore a simple Gaussian does not precisely match the observed eclipse profile.  In addition, our analytic model cannot reproduce light curves with extreme ellipsoidal modulations, i.e. systems with tidally deformed and oblate stars.  Nonetheless, our analytic model captures the basic light curve parameters of eclipse depths, eclipse widths, eclipse phases, and amplitude of the reflection effect.  These parameters are sufficient in allowing us to distinguish different classes of eclipsing binaries.

\begin{figure*}[t]
\centerline{
\includegraphics[trim= 0.0cm 0.0cm 0.1cm 0.1cm, clip=true, width=6.7in]{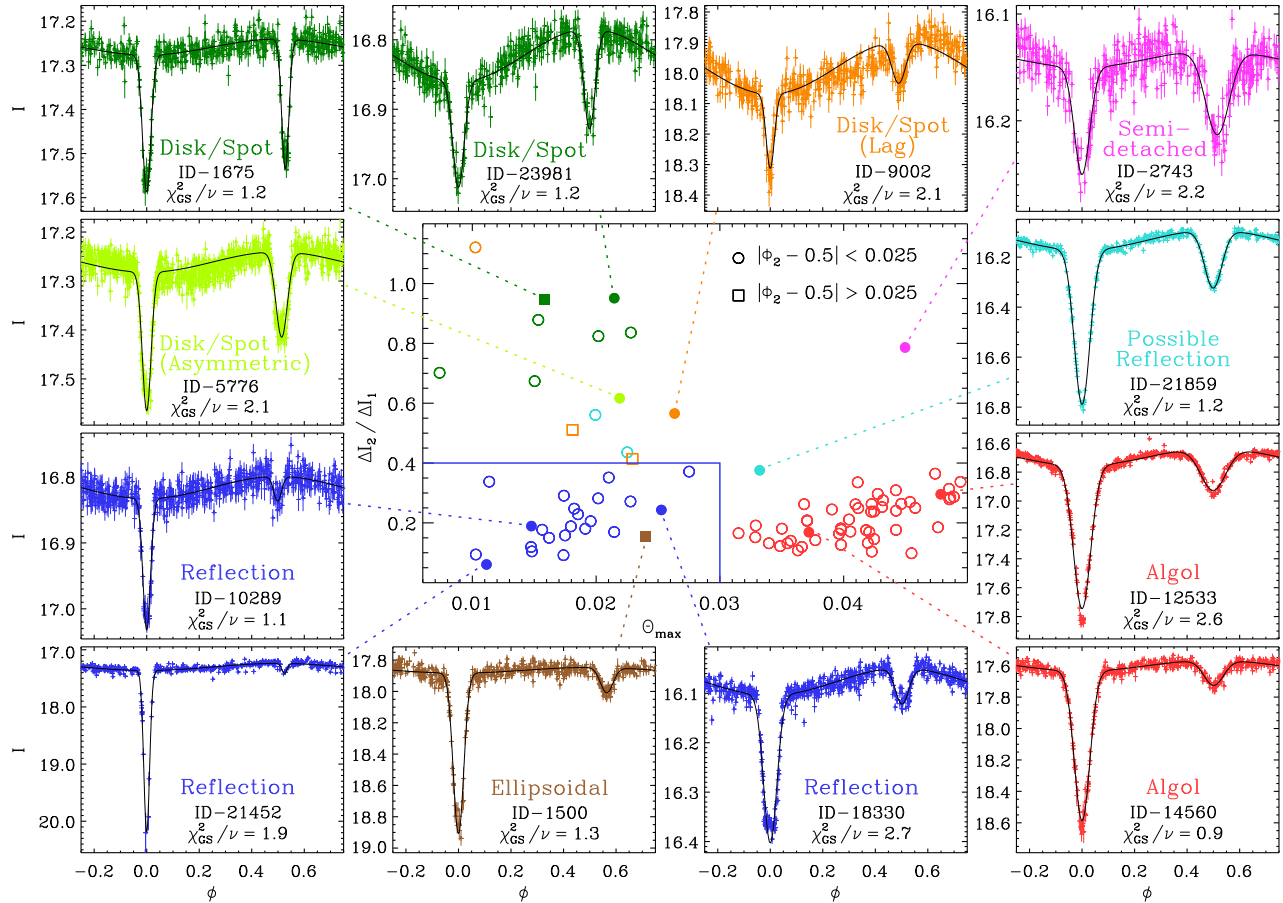}}
\caption{Outside panels: Example light curves of the various eclipsing binary populations and the best {\it analytic} fits of Gaussians and sinusoids.   Center panel: Ratio of eclipse depths $\Delta I_2/\Delta I_1$ versus maximum eclipse width $\Theta_{\rm max}$ = max($\Theta_1,\Theta_2)$ for the 90 eclipsing binaries with B-type MS primaries, $P$ = 3\,-\,15 days, $\Delta I_{\rm refl}$ $>$ 0.015 mag, and $\Theta_{\rm max}$ $<$ 0.05.  Eclipsing binaries with components of comparable luminosity are toward the top and those with a component that fills or nearly fills its Roche lobe are toward the right.  We also distinguish systems with eccentric orbits $e$ $>$ $e_{\rm min}$ $>$ 0.04 (square symbols) from those that most likely have nearly circular orbits (circles) based on the observed orbital phase of the secondary eclipse $\Phi_2$.  Dotted lines and filled symbols match each light curve to the corresponding system in the central panel.  A semi-detached binary (magenta) is to the upper right. Evolved semi-detached Algols (red) with broad eclipse features, large temperature contrasts, and inverted mass ratios are to the bottom right.   Eclipsing binaries with excess luminosity from an accretion disk and/or hot spot (green) are to the upper left, including one system with an asymmetric light curve profile (light green) and four systems with a peak in the light curve that lags the secondary eclipse (orange).   One system exhibits ellipsoidal modulation (brown) in a moderately eccentric orbit.  The 22 eclipsing binaries exhibiting genuine reflection effects (blue) have nearly circular orbits and form a distinct population toward the bottom left with $\Theta_{\rm max}$~$<$~0.03 and $\Delta I_2$/$\Delta I_1$ $<$ 0.4 (blue solid lines).  The three systems that may also show reflection effects (cyan) have slightly different properties and lie outside our adopted parameter space. }
\end{figure*}

To identify eclipsing binaries with reflection effects and well-defined eclipses, we impose the following selection criteria.  We require the reflection effect amplitude to be $\Delta I_{\rm refl}$ $>$ 0.015 mag and its 1$\sigma$ uncertainty to be $<$20\% of its value.  We stipulate that the 1$\sigma$ uncertainties in the eclipse depths $\Delta I_1$ and $\Delta I_2$ and eclipse widths $\Theta_1$ and $\Theta_2$ are $<$25\% their respective values.  We discard eclipsing binaries with wide eclipses $\Theta_{\rm max}$~=~max($\Theta_1,\Theta_2)$~$>$~0.05, which removes most systems that have filled their Roche lobes, e.g. semi-detached and contact binaries. Eclipsing binaries with shallow eclipse depths can remain undetected given the sensitivity and cadence of the OGLE-III LMC observations.  We therefore keep only systems with total light curve amplitudes:

\begin{equation}
 \Delta I   = \Delta I_1 + \Delta I_{\rm refl} \ge 10^{0.2(\langle I \rangle-16.0)} \times 0.08\,{\rm mag}
\end{equation}

\noindent to ensure our sample is complete in our selected parameter space (see detection limits in Fig.~3 of \citealt{Graczyk2011}).

\subsection{Results}

We find 90 eclipsing binaries that satisfy these initial selection criteria (see Fig.~3).  In this subsample, there is one semi-detached binary (magenta system in Fig.~3) and 51 Algols, i.e. evolved semi-detached eclipsing binaries that have inverted their mass ratios via stable mass transfer (red population in Fig.~3).  These evolved eclipsing binaries have wide eclipses 0.031~$<$~$\Theta_{\rm max}$~$<$~0.050 that dictate at least one of the binary components fills their Roche lobe. Previous studies of eclipsing binaries in the LMC have noted this Algol population by identifying systems with wide eclipses and large temperature contrasts \citep{Mazeh2006,Prsa2008}. 

The remaining 38 objects that satisfy our initial selection criteria have intriguing light curves.  We list their catalog properties and analytic light curve parameters in Table~1. All but one of these eclipsing binaries have $\Theta_{\rm max}$ $<$ 0.028, which indicate a detached configuration.   In the following, we use the measured analytic parameters of these 38 systems to understand their physical properties as well as to distinguish various classes of eclipsing binaries. 

\subsubsection{MS\,+\,Pre-MS ``Reflecting'' Eclipsing Binaries \\ with Extreme Mass Ratios}

Of the 38 unusual eclipsing binaries, we discover that 22 systems form a distinct population with definitive reflection effects $\Delta I_{\rm refl}$ $>$ 0.015 mag, narrow eclipses $\Theta_{\rm max}$ $<$ 0.03, relatively shallow secondary eclipses $\Delta I_2$/$\Delta I_1$ $<$ 0.4, and nearly circular orbits according to $\Phi_2$  and Eq. 2 (blue systems in Fig.~3).  These 22 eclipsing binaries have short orbital periods $P$~=~3.0\,-\,8.5 days, large reflection effect amplitudes  $\Delta I_{\rm refl}$ = 0.017\,-\,0.138~mag, and moderate to deep primary eclipses $\Delta I_1$~=~0.09\,-\,2.8~mag.  

The reflection effects and primary eclipse depths can be so prominent only if the companions are comparable in size to but substantially cooler than the B-type MS primaries. The companions cannot be normal MS stars since cooler MS stars are also considerably smaller.  We can eliminate the alternative that the companions are evolved cool subgiants in an Algol binary because such large subgiants fill their Roche lobes and produce markedly wider eclipses.   We therefore conclude that the companions in our 22 systems are cool medium-sized low-mass pre-MS stars that have not yet fully formed. 

We can observe these nascent B-type MS + low-mass pre-MS eclipsing binaries at such a special time in their evolution because low-mass companions $q$~$\lesssim$~0.25 contract considerably more slowly during their pre-MS phase of formation.  See \S3, where we more thoroughly analyze the physical properties of these systems by fitting detailed physical light curve models.  Our 22 eclipsing binaries with pronounced reflection effects therefore constitute a new class of detached MS\,+\,pre-MS close binaries with extreme mass ratios. These systems also represent the first unambiguous identification of B-type MS stars with closely orbiting low-mass stellar companions.

In addition to the selection effects discussed in \S1, the difficulty in detecting low-mass eclipsing companions to B-type MS stars partially stems from the small number of nearby B-type MS stars in our Milky Way galaxy.  Quantitatively, there are only $\approx$6,000 B-type stars  with robust parallactic distances $d$~$<$~500\,pc \citep{Perryman1997}.  This is a factor of $\approx$30 times smaller than the number of B-type MS stars ${\cal N}_{\rm B}$~$\approx$~174,000 in our OGLE-III LMC sample.  It is therefore not surprising that we have not yet observed in the Milky Way the precise counterparts to our reflecting eclipsing binaries with B-type MS primaries and low-mass pre-MS companions. 

\begin{table*}[t]\scriptsize
\begin{flushleft}
{\small {\bf Table 1}. Analytic model parameters of 38 eclipsing binaries with intriguing light curves.}
\end{flushleft}
\renewcommand{\tabcolsep}{4.6pt}
\vspace{-0.3cm}
\begin{tabular}{|r|r|c|r|c|c|c|c|c|c|c|c|c|c|l|}
\hline
\multicolumn{3}{|c|}{Catalog Properties} & \multicolumn{12}{c|}{Analytic Model Parameters}\\
\hline
     ID~~  & $\langle V-I \rangle$ & ${\cal N}_I$ & $P$~~~~ & $t_{\rm o}$ & 
     $\langle I \rangle$ & $\Delta I_1$ & $\Theta_1$ & $\Phi_2$ & $\Delta I_2$ & $\Theta_2$ &
     $\Delta I_{\rm refl}$ & ${\cal N}_{\rm c}$ & $\chi^2_{\rm GS}/\nu$ & Type \\
    \hline
    1500 &    0.17~ & 440 & 6.025412 & 3563.908 & 17.883 & 1.023 & 0.0201 & 0.565 & 0.160 & 0.0240 & 0.035 & 1 & 1.29 & Ellipsoidal \\
    \hline
    1675 &    0.04~ & 465 & 8.697351 & 3568.781 & 17.277 & 0.310 & 0.0159 & 0.528 & 0.294 & 0.0136 & 0.037 & 1 & 1.21 & Disk/Spot \\
    \hline
    1803 & $-$0.07~ & 460 & 3.970086 & 3572.117 & 17.090 & 0.637 & 0.0162 & 0.501 & 0.096 & 0.0156 & 0.138 & 1 & 1.55 & Reflection \\
    \hline
    1965 & $-$0.09~ & 440 & 3.175248 & 3570.628 & 17.221 & 0.197 & 0.0219 & 0.497 & 0.073 & 0.0275 & 0.017 & 2 & 1.10 & Reflection \\
    \hline
    2139 & $-$0.16~ & 477 & 8.462510 & 3576.045 & 16.507 & 0.328 & 0.0103 & 0.502 & 0.031 & 0.0089 & 0.033 & 2 & 1.28 & Reflection \\
    \hline
    3972 &    0.06~ & 448 & 3.396033 & 3565.053 & 17.882 & 0.263 & 0.0170 & 0.496 & 0.148 & 0.0199 & 0.023 & 2 & 4.14 & Possible Reflection \\
    \hline
    5205 & $-$0.04~ & 456 & 4.164725 & 3564.819 & 16.682 & 0.162 & 0.0225 & 0.499 & 0.071 & 0.0221 & 0.044 & 1 & 1.22 & Possible Reflection \\
    \hline
    5377 & $-$0.11~ & 447 & 3.276364 & 3563.150 & 17.346 & 0.165 & 0.0178 & 0.502 & 0.047 & 0.0202 & 0.020 & 1 & 1.18 & Reflection \\
    \hline
    5776 & $-$0.03~ & 856 & 4.715550 & 3564.833 & 17.282 & 0.283 & 0.0166 & 0.514 & 0.174 & 0.0219 & 0.042 & 2 & 2.06 & Disk/Spot (Asym.) \\
    \hline
    5898 & $-$0.05~ & 439 & 5.323879 & 3567.545 & 16.752 & 0.834 & 0.0212 & 0.501 & 0.141 & 0.0215 & 0.098 & 1 & 1.13 & Reflection \\
    \hline
    6630 & $-$0.02~ & 410 & 3.105571 & 3563.821 & 17.104 & 0.164 & 0.0205 & 0.497 & 0.058 & 0.0210 & 0.019 & 1 & 1.16 & Reflection \\
    \hline
    7419 &    0.13~ & 421 & 4.255889 & 3563.579 & 16.502 & 0.185 & 0.0180 & 0.497 & 0.035 & 0.0135 & 0.037 & 2 & 1.11 & Reflection \\
    \hline
    7842 & $-$0.05~ & 477 & 3.781798 & 3565.825 & 17.966 & 1.725 & 0.0148 & 0.498 & 0.181 & 0.0148 & 0.083 & 1 & 1.24 & Reflection \\
    \hline
    9002 &    0.04~ & 424 & 3.578291 & 3568.696 & 18.069 & 0.243 & 0.0156 & 0.490 & 0.137 & 0.0263 & 0.173 & 2 & 2.08 & Disk/Spot (Lag) \\
    \hline
    9642 &    0.11~ & 782 & 3.913360 & 3565.613 & 17.933 & 0.699 & 0.0174 & 0.502 & 0.064 & 0.0143 & 0.050 & 1 & 1.23 & Reflection \\
    \hline
   10289 & $-$0.09~ & 557 & 4.642567 & 3566.031 & 16.833 & 0.199 & 0.0144 & 0.499 & 0.038 & 0.0148 & 0.034 & 1 & 1.09 & Reflection \\
    \hline
   10941 & $-$0.02~ & 559 & 4.079727 & 3563.840 & 17.585 & 0.154 & 0.0211 & 0.401 & 0.064 & 0.0229 & 0.041 & 2 & 1.26 & Disk/Spot (Lag) \\
    \hline
   11731 & $-$0.04~ & 477 & 5.661823 & 3544.866 & 16.837 & 0.059 & 0.0181 & 0.529 & 0.030 & 0.0169 & 0.065 & 1 & 1.65 & Disk/Spot (Lag) \\
    \hline
   11787 & $-$0.03~ & 493 & 3.305829 & 3542.971 & 17.757 & 0.487 & 0.0202 & 0.500 & 0.401 & 0.0201 & 0.024 & 0 & 1.09 & Disk/Spot \\
    \hline
   12528 &    0.03~ & 476 & 8.187359 & 3537.820 & 17.749 & 0.240 & 0.0153 & 0.498 & 0.211 & 0.0153 & 0.019 & 2 & 1.15 & Disk/Spot \\
    \hline
   13194 &    0.05~ & 493 & 11.535650 & 3557.945 & 17.826 & 0.234 & 0.0079 & 0.498 & 0.262 & 0.0103 & 0.048 & 2 & 1.36 & Disk/Spot (Lag) \\
    \hline
   13721 & $-$0.10~ & 428 & 3.122558 & 3558.192 & 17.818 & 0.420 & 0.0169 & 0.500 & 0.096 & 0.0185 & 0.025 & 1 & 1.01 & Reflection \\
    \hline
   15306 & $-$0.03~ & 537 & 12.654262 & 3580.455 & 17.955 & 0.612 & 0.0068 & 0.501 & 0.430 & 0.0074 & 0.058 & 2 & 0.92 & Disk/Spot \\
    \hline
   15761 & $-$0.11~ & 223 & 5.310911 & 3581.568 & 16.530 & 0.850 & 0.0147 & 0.504 & 0.101 & 0.0142 & 0.108 & 2 & 2.86 & Reflection \\
    \hline
   15792 & $-$0.11~ & 600 & 4.317022 & 3566.098 & 16.721 & 0.231 & 0.0140 & 0.500 & 0.041 & 0.0156 & 0.039 & 2 & 1.63 & Reflection \\
    \hline
   16828 & $-$0.16~ & 606 & 3.675697 & 3572.011 & 16.618 & 0.109 & 0.0174 & 0.502 & 0.032 & 0.0146 & 0.018 & 2 & 1.26 & Reflection \\
    \hline
   17217 & $-$0.12~ & 592 & 5.354795 & 3576.802 & 16.543 & 0.098 & 0.0114 & 0.490 & 0.033 & 0.0096 & 0.022 & 2 & 1.24 & Reflection \\
    \hline
   17387 & $-$0.17~ & 605 & 4.772926 & 3567.419 & 16.157 & 0.094 & 0.0219 & 0.499 & 0.026 & 0.0228 & 0.017 & 0 & 1.15 & Reflection \\
    \hline
   17695 &    0.01~ & 473 & 3.096068 & 3567.092 & 17.343 & 0.176 & 0.0208 & 0.501 & 0.148 & 0.0228 & 0.019 & 2 & 0.92 & Disk/Spot \\
    \hline
   18330 & $-$0.01~ & 599 & 3.252913 & 3561.424 & 16.104 & 0.297 & 0.0253 & 0.502 & 0.073 & 0.0252 & 0.056 & 2 & 2.75 & Reflection \\
    \hline
   18419 & $-$0.13~ & 599 & 4.118145 & 3564.319 & 16.741 & 0.545 & 0.0158 & 0.485 & 0.086 & 0.0175 & 0.059 & 2 & 1.83 & Reflection \\
    \hline
   19186 &    0.01~ & 473 & 5.652474 & 3567.458 & 17.701 & 0.278 & 0.0130 & 0.500 & 0.187 & 0.0150 & 0.019 & 1 & 0.94 & Disk/Spot \\
    \hline
   21025 & $-$0.10~ & 435 & 4.543290 & 3581.225 & 16.777 & 0.134 & 0.0182 & 0.491 & 0.033 & 0.0138 & 0.025 & 1 & 0.97 & Reflection \\
    \hline
   21452 &    0.17~ & 377 & 8.178958 & 3541.898 & 17.362 & 2.824 & 0.0111 & 0.525 & 0.172 & 0.0098 & 0.124 & 2 & 1.86 & Reflection \\
    \hline
   21641 &    0.05~ & 436 & 3.092424 & 3572.506 & 16.741 & 0.405 & 0.0191 & 0.500 & 0.073 & 0.0191 & 0.062 & 1 & 0.76 & Reflection \\
    \hline
   21859 &    0.00~ & 428 & 3.154536 & 3571.735 & 16.172 & 0.617 & 0.0296 & 0.499 & 0.232 & 0.0332 & 0.080 & 1 & 1.23 & Possible Reflection \\
    \hline
   21975 &    0.13~ & 437 & 3.021139 & 3571.601 & 16.208 & 0.139 & 0.0196 & 0.500 & 0.029 & 0.0160 & 0.030 & 0 & 1.11 & Reflection \\
    \hline
   23981 &    0.09~ & 435 & 5.720508 & 3579.130 & 16.861 & 0.152 & 0.0215 & 0.499 & 0.144 & 0.0213 & 0.076 & 0 & 1.17 & Disk/Spot \\
    \hline
    
\end{tabular}
\end{table*}

\subsubsection{Other Intriguing Light Curves}

We now discuss the remaining 16 unusual systems in our OGLE-III LMC sample.  The properties of these 16 eclipsing binaries are not fully understood, and may potentially have important implications for the evolution of close binaries.    However, they have distinctly different light curve parameters and physical characteristics than those in our 22 reflecting MS\,+\,pre-MS eclipsing binaries.  A detailed study of these 16 unusual systems is therefore not in the scope of the present study. We only summarize the observed properties of these 16 systems to illustrate the uniqueness of our nascent eclipsing binaries.  

 The 12 eclipsing binaries in the top left of Fig.~3 have deep secondary eclipses and large out-of-eclipse variations that are not necessarily symmetric with respect to the eclipses. The lack of symmetry dictates that the variations cannot be solely due to reflection effects.  Moreover, the deeper secondary eclipses in these systems indicate excess light from a hot spot and/or accretion disk. Similar systems exist in our Milky Way such as V11 in the old open cluster NGC 6791 \citep{deMarchi2007}, T-And0-00920 in the galactic field \citep{Devor2008}, and SRa01a\_34263 in the young open cluster NGC 2264 \citep{Klagyivik2013}.  Quantitatively, these 12 eclipsing binaries with luminous disks and/or hot spots have $\Delta I_2$/$\Delta I_1$ $>$ 0.4, while our 22 systems with low-luminosity pre-MS companions have $\Delta I_2$/$\Delta I_1$ $<$ 0.4.  

Our 22 eclipsing binaries with pre-MS companions have nearly circular orbits with $| \Phi_2 - \nicefrac{1}{2} |$ $\le$ 0.025, as expected from tidal damping even earlier in their pre-MS phase of evolution \citep{Zahn1989}. One peculiar eclipsing binary, ID-1500 (brown system in Fig.~3), satisfies our selection criteria of $\Theta_{\rm max}$~$<$~0.03 and $\Delta I_2$/$\Delta I_1$ $<$ 0.4, but has a moderately eccentric orbit of $e$ $>$ $e_{\rm min}$($\Phi_2$\,=\,0.565) = 0.10 according to Eq.~2.    This eclipsing binary has a deep primary eclipse $\Delta I_1$~=~1.0~mag that stipulates the binary components cannot both be normal MS stars.  However, the light curve of ID-1500 peaks at $\phi$ $\approx$ 0.8, i.e. $\phi$ $\approx$ $-$0.2 as folded in Fig.~3, suggesting the out-of-eclipse variations are due to ellipsoidal modulations in an eccentric orbit instead of reflection effects.  Specifically, periastron in this system probably occurs near $\phi$ $\approx$ $-$0.2, at which point the stars are tidally deformed into oblate ellipsoids and the perceived flux is increased.   We attempt to fit a detailed physical model (see \S3) to this system assuming the companion is a pre-MS star, but our fit is rather poor with $\chi^2/\nu$ = 1.7.  Moreover, our physical model converges toward an unrealistic solution with $q$ $\approx$ 0.5.  Whether the out-of-eclipse variations in this system are due to reflection effects or are entirely because of ellipsoidal modulations, the removal of this one system does not affect our investigation of low-mass $q$ $<$ 0.25 companions to B-type MS stars.  

We find three additional eclipsing binaries that may display reflection effects with a pre-MS companion, but lie just outside of our selected parameter space (cyan systems in Fig.~3).  ID-21859 has a broad eclipse $\Theta_{\rm max}$~=~0.033, but has eclipse depth properties that separate it from the observed Algol population. ID-3972 and ID-5205 have slightly deeper secondary eclipses $\Delta I_2$/$\Delta I_1$ $\approx$ 0.5, but exhibit symmetric light curve profiles with no immediate indications of disks and/or hot spots.  In our Monte Carlo simulations (\S5), we implement the same selection criteria utilized here, and so we do not include these three systems in our statistical sample.   Moreover, the observed population of eclipsing binaries with genuine reflection effects are concentrated near $\Theta_{\rm max}$ $\approx$ 0.018 and $\Delta I_2$/$\Delta I_1$~$\approx$~0.2.  Increasing distance from this center according to our adopted metric increases the likelihood that the system is not a MS\,+\,pre-MS eclipsing binary.  Our 22 eclipsing binaries that exhibit pronounced reflection effects $\Delta I_{\rm refl}$~$>$~0.015~mag with pre-MS companions at $P$ $=$ 3.0\,-\,8.5 days have $\Theta_{\rm max}$ $\le$ 0.03, which cleanly differentiates them from Algols and contact binaries that fill their Roche lobes, $\Delta I_2$/$\Delta I_1$ $\le$ 0.4, which distinguishes them from systems with luminous disks and/or hot spots, and $|\Phi_2 - \nicefrac{1}{2}|$ $\le$ 0.025, which separates them from systems that show ellipsoidal modulations in an eccentric orbit (Fig.~3). We emphasize that these criteria are rather effective in selecting systems with low-mass pre-MS companions while simultaneously minimizing contamination from other types of eclipsing binaries.

\subsection{Comparison to Previously Known Classes}

\subsubsection{Irradiated Binaries}

Other classes of detached binaries can exhibit intense irradiation effects,  but there are key differences that distinguish our 22 systems.  Namely, our 22 eclipsing binaries contain a hot MS primary with a cool pre-MS companion, while most previously known reflecting eclipsing binaries contain a hot evolved remnant with a cool MS companion  \citep{Bond2000,Lee2009}.  For example,  eclipsing binaries with subdwarf B-type (sdB) primaries and M-dwarf companions, sometimes called HW Vir eclipsing binaries after the prototype, have similar reflection effect amplitudes and light curve properties \citep{Lee2009,Barlow2013,Pietrukowicz2013}.  However, HW Vir systems differ from our systems in three fundamental parameters.  First, the sdB primaries in HW Vir eclipsing binaries are intrinsically $\sim$100 times less luminous than B-type MS stars, and would therefore not be detectable in the LMC given the sensitivity of the OGLE-III survey. Second,  HW Vir eclipsing binaries have shorter orbital periods $P$~$\lesssim$~0.5~days than our 22 systems with $P$~=~3.0\,-\,8.5~days.  This is because the sdB primaries and M-type MS secondaries are smaller and less luminous than our B-type MS primaries and pre-MS companions, and therefore must be closer together to produce observable reflection effects.  Finally, HW~Vir systems are evolved binaries and associated with old stellar populations, while our 22 nascent eclipsing binaries are situated in or near star-forming H\,\textsc{ii} regions (see \S4).  

As another example, binaries in which a MS star orbits the hot central star of a planetary nebula can pass through a very brief interval $\lesssim$10,000\,yrs when reflection effects are detectable, although eclipses are generally not observed \citep{Bond2000}. Such systems could have satisfied our magnitude and color criteria, but these binaries are typically at shorter periods $P$~$<$~3~days than we have selected \citep{Miszalski2009}. Moreover, we cross-referenced the positions of our 22 systems with catalogs of planetary nebulae \citep{Reid2010}  and emission-line point sources \citep{Howarth2013} in the LMC, and do not find any matches. Our nascent B-type MS + pre-MS eclipsing binaries clearly exhibit a phenomenologically different type of reflection effect than those observed in evolved binaries with stellar remnants.

\subsubsection{Pre-MS Binaries}

Although there is a rich literature regarding pre-MS binaries (see \citealt{Hillenbrand2004} and review by \citealt{Mathieu1994}), only a few close MS\,+\,pre-MS binaries have been identified. For example, photometric and spectroscopic observations of the eclipsing binaries EK~Cep \citep{Popper1987}, AR~Aur \citep{Nordstrom1994}, TY~CrA \citep{Casey1998}, and RS~Cha \citep{Alecian2007} have demonstrated the primaries are close to the zero-age MS while the secondaries are still contracting on the pre-MS.   However, these systems have late-B/A-type MS primaries ($M_1$~$\approx$~1.9\,-\,3.2\,\Msun), components of comparable mass ($q$ $\approx$ 0.5\,-\,1.0), and temperature contrasts ($T_2$/$T_1$ $\approx$ 0.4\,-\,0.9) that are too small to produce detectable reflection effects.  \citet{MoralesCalderon2012} identified ISOY~J0535-447 as a young pre-MS eclipsing binary with an extreme mass ratio $q$~$\approx$~0.06, but with a low-mass $M_1$ $\approx$ 0.8\,\Msun\ early-K primary.  

The only similar analog of a B-type MS primary with a closely orbiting low-mass pre-MS companion is the eclipsing binary BM~Orionis \citep{Hall1969,Palla2001,Windemuth2013}, although the nature of its secondary has been debated and has even been suggested to be a black hole \citep{Wilson1972}.  Located in the heart of the Orion Nebula, BM Ori exhibits broad eclipses with noticeable undulations in the eclipse shoulders. These features indicate the companion nearly fills its Roche lobe and is still accreting from the surrounding disk. If BM Ori contains an accreting pre-MS companion, then it could be a precursor to our 22 eclipsing binaries that show no evidence for an accretion disk. Indeed, BM Ori is extremely young with an age $\tau$~$\lesssim$~0.1~Myr estimated from pre-MS contraction timescales \citep{Palla2001} and the dynamics of the inner region of the Orion Nebula \citep{ODell2009}.  Meanwhile, the disk photoevaporation timescale around Herbig Be pre-MS stars with $M_1$ $\approx$ 3\,-\,8\,\Msun\ is $\approx$0.3 Myr \citep{AlonsoAlbi2009}.  It is therefore not unexpected that BM Ori at $\tau$~$<$~0.1~Myr still has a disk.  Alternatively, our 22 systems no longer have a noticeable accretion disk in the photometric light curves, and so must be older than $\tau$ $\gtrsim$ 0.3 Myr (see also \S3.3).  

If BM Ori was placed in the LMC and observed by the OGLE-III survey, it would not be contained in our sample for three reasons.  First, BM Ori contains an extremely young and reddened mid-B MS primary with $M_1$ $\approx$ 6\,\Msun, $\langle V - I \rangle$ $\approx$ 0.8, and  $\langle I \rangle$ $\approx$ 8.8 at the distance $d$ $\approx$ 400 pc to the Orion Nebula \citep{Windemuth2013}.  It would be rather faint at $\langle I \rangle$ $\approx$ 19.3 if located at the distance $d$ = 50 kpc to the LMC, and therefore below our photometric selection limit.  Second, even if we extended our search toward fainter systems, the reflection effect amplitude in BM~Ori is too small to be observed given the sensitivity of the OGLE-III LMC observations. Finally, the secondary eclipse is extremely shallow with undulations in the eclipse shoulders.  We could not measure well-defined secondary eclipse parameters according to our analytic model.  In addition to being at a fundamentally different stage of evolution, i.e. still accreting from a disk,  BM Ori has clearly different photometric light curve properties than those of our 22 eclipsing binaries. 

Finally, BM Ori has a modest mass ratio $q$~=~0.31.  In contrast, the majority of our reflecting eclipsing binaries have extreme mass ratios $q$ $<$ 0.25 (see below), as indicated by their more luminous, massive primaries and larger reflection effect amplitudes. Our reflecting eclipsing binaries represent the first detection of B-type MS stars with close extreme mass-ratio companions where the orbital periods and the nature of the companions are reliably known.

\section{Physical Properties}

\subsection{Overview of Methodology}

The component masses in eclipsing binaries are typically measured dynamically via spectral radial velocity variations.  However, our eclipsing binaries in the LMC are relatively faint 16 $<$ $\langle I \rangle$ $<$ 18 and typically embedded in H\,\textsc{ii} regions (\S4) that would contaminate the stellar spectra with nebular emission lines.  Moreover, B-type MS stars experience slight atmospheric variations and rotate so rapidly that their spectral absorption lines are generally broadened by $v_{\rm surface}$ $\approx$ 100\,-\,250~km~s$^{-1}$ \citep{Abt2002,Levato2013}. There is a small population of slowly rotating B-type MS stars with $v_{\rm surface}$ $\approx$ 50~km~s$^{-1}$, and \citet{Abt2002} and \citet{Levato2013} suggest these systems may by tidally synchronized with closely orbiting low-mass companions.  Indeed, our reflecting eclipsing binaries may partially explain the origins of B-type MS slow rotators.  In any case,  it would be quite observationally expensive to detect small velocity semi-amplitudes $K_1$~$\approx$~25\,($q$/0.1)\,km~s$^{-1}$ induced by closely orbiting low-mass companions for all 22 eclipsing binaries in our statistical sample.  In the future, we plan to obtain multi-epoch spectra for a small subset of our MS\,+\,pre-MS eclipsing binaries.   To analyze all 22 systems, however,  we must currently utilize a different technique of inferring the physical properties based solely on the observed photometric light curves.

 Fortunately, we have two additional constraints that allow us to estimate the masses of the binary components from the observed eclipse properties.  First, our eclipsing binaries are detached from their Roche lobes, as demonstrated by their narrow eclipses, and therefore the primary and secondary are each effectively evolving along their respective single-star sequences (see \S3.4 for further justification of this expectation and a discussion of systematic uncertainties).  Given an age $\tau$ and masses $M_1$ and $M_2$, we can interpolate stellar radii $R_1$ and $R_2$, photospheric temperatures $T_1$ and $T_2$, and luminosities $L_1$ and $L_2$ from theoretical stellar evolutionary tracks.  We can then use empirical bolometric corrections and color indices to map the physical properties of the eclipsing binaries into observed magnitudes and colors.  \citet{Devor2006} and \citet{Devor2008} employed a similar technique of estimating ages and masses of galactic eclipsing binaries by incorporating stellar isochrones into their photometric light curve modeling.  Their algorithm worked for a small subset of systems.  In general, however, the parameters $\tau$, $M_1$, and $M_2$ were generally degenerate, not unique, and/or not constrained.

This brings us to our second constraint.  Unlike the sample of galactic eclipsing binaries studied by \citet{Devor2008}, we know the distances to our 22 eclipsing binaries in the LMC.  This extra distance constraint fully eliminates the degeneracy and allows us to calculate unique solutions for the physical properties of the eclipsing binaries.  The deductions of the physical parameters progress as follows.  The measured mean magnitude $\langle I \rangle$ and color $\langle V - I \rangle$, along with the distance, bolometric corrections, and color indices, mainly provide the luminosity $L_1$ of the B-type MS primary and the amount of dust reddening $E(V-I)$, respectively.  From MS stellar evolutionary tracks, we can estimate the mass $M_1$ and radius $R_1$ of a young B-type MS star with luminosity $L_1$. The amplitude of the reflection effect $\Delta I_{\rm refl}$ is an indicator of $T_2$/$T_1$ and $R_2$/$R_1$ as discussed in \S2, but also depends on the albedo of the secondary $A_2$.  The sum of eclipse widths $\Theta_1 + \Theta_2$ determines the sum of the relative radii $(R_1 + R_2)/a$, the ratio of eclipse depths $\Delta I_2 / \Delta I_1$ gives the luminosity contrast $L_2$/$L_1$, and the magnitude of the primary eclipse depth $\Delta I_1$ provides the inclination $i$.   Since we already know $M_1$, $R_1$ and $L_1$, we can infer $R_2$ and $L_2$ directly from the observed light curve parameters.  Finally, according to pre-MS evolutionary tracks, the radius $R_2$ and luminosity $L_2$ of the pre-MS secondary uniquely corresponds to its age $\tau$ and mass $M_2$.  In our full procedure (see below), we calculate each of these parameters simultaneously in a self-consistent manner.  We also consider various sources of systematic errors in our measured light curve parameters as well as stellar evolutionary tracks. Nonetheless, the steps discussed above illustrate how we can estimate the physical properties of detached, unevolved, eclipsing binaries with known distances using only the photometric light curves.

\subsection{Physical Model Fits}

   In our eclipsing binary models, we have eight physical parameters: orbital period $P$, epoch of primary eclipse minimum $t_{\rm o}$, primary mass $M_1$, secondary mass $M_2$, age $\tau$, inclination $i$, albedo of the secondary $A_2$, and amount of dust extinction $A_I$ toward the system.    B-type MS stars in the LMC have slightly subsolar metallicities log($Z$/\Zsun)~$\approx$~$-0.4$ \citep{Korn2000}, where \Zsun~$\approx$~0.015. We therefore incorporate the Padova Z=0.008, Y=0.26 stellar evolutionary tracks to describe the MS evolution \citep{Bertelli2009}, and the Pisa Z=0.008, Y=0.265, $\alpha$ = 1.68, $X_D$=2$\times$10$^{-5}$ tracks to model the pre-MS evolution \citep{Tognelli2011}.  The physical properties of the binary components, e.g. radii $R_1$ and $R_2$, surface temperatures $T_1$ and $T_2$, luminosities $L_1$ and $L_2$, and surface gravities $g_1$ and $g_2$,  are then interpolated from these stellar tracks according to the model parameters $M_1$, $M_2$, and $\tau$.  We use updated, temperature-dependent color indices and bolometric corrections \citep{Pecaut2013} to transform the intrinsic luminosities and temperatures of both binary components into combined absolute magnitudes $M_I$ and $M_V$.  We adopt the dust reddening law of $E(V$\,$-$\,$I)$ = 0.7$A_I$ \citep{Cardelli1989, Fitzpatrick1999,Ngeow2005} and LMC distance modulus of $\mu$ = 18.5 \citep{Pietrzynski2013} to then calculate the observed magnitudes $\langle I \rangle$~=~$M_I$~+~$\mu$~+~$A_I$ and $\langle V \rangle$~=~$M_V$~+~$\mu$~+~1.7$A_I$. 

We~primarily~utilize~the~eclipsing~binary~modeling software~\textsc{Nightfall}\footnote{http://www.hs.uni-hamburg.de/DE/Ins/Per/Wichmann/ \\~~~Nightfall.html}~to~synthesize~$I$-band~and~$V$-band light~curves.~We~implement~a~square-root~limb~darkening law~with~the~default~limb-darkening~coefficients,~the default~gravity~brightening~coefficients,~model~atmospheres according~to~the~surface~gravities~of~the~binary components,~fractional~visibility~of~surface~elements, three~iterations~of~reflection~effects,~and~the~default albedo~of~$A_1$\,=\,1.0~for~the~hot~B-type~MS~primaries.  

Given the sensitivity of the OGLE-III data, the 19 eclipsing binaries with $|\Phi_2 - \nicefrac{1}{2}|$ $<$ 0.01 and $\Theta_1$~$\approx$~$\Theta_2$ have $e$~$<$~0.02 (see Eqn.~2 and \citealt{Kallrath2009}).  For these 19 systems, we assume  circular orbits in our physical models.  The three systems (ID-17217, ID-18419, and ID-21452) in slightly eccentric orbits with  0.010~$\le$~$|\Phi_2$\,$-$\,$\nicefrac{1}{2}|$~$\le$~0.025  have longer orbital periods where tidal effects are not as significant.   The eclipse widths $\Theta_1$~$\approx$~$\Theta_2$ are also comparable to each other in these three eclipsing binaries, dictating the eccentricities 0.02 $<$ $e$ $<$ 0.08 are small.  Because the orbits are so close to circular, we cannot easily break the degeneracy between the eccentricity $e$ and the argument of periastron $\omega$.  For these three systems, we impose $e$~=~$e_{\rm min}$/$\langle$cos($\omega$)$\rangle$~=~1.6\,$e_{\rm min}$ according to Eqn. 2, where we have assumed a uniform probability distribution for $\omega$.  Adjusting the eccentricities to values within  $e_{\rm min}$~$<$~$e$~$<$~2.2$e_{\rm min}$ do not change the fitted model parameters beyond the uncertainties.   For ID-21452, which has $\Phi_2$ $>$ 0.5, we assume $\omega$ = 50$^{\rm o}$.  For ID-17217 and ID-18419, which have $\phi_2$ $<$ 0.5, we adopt $\omega$ = 230$^{\rm o}$.  Changing the argument of periastron to the opposite angle, e.g. $\omega$ = 310$^{\rm o}$ for ID-21452 or $\omega$ = 130$^{\rm o}$ for ID-17217 and ID-18419, has a negligible effect on the other model parameters considering the eccentricities are so small. 

 Since tides have fully or nearly circularized the orbits, the rotation rates of the pre-MS companions with large convective envelopes are expected to be tidally synchronized with the orbital periods \citep{Zahn1989}.   For example, a 1.5\,\Msun\ pre-MS star with age $\tau$~=~1~Myr in a $P$~=~4~day orbit with a 10\,\Msun\ B-type MS star has rapid synchronization and spin-orbit alignment timescales of $\lesssim$\,0.01~Myr \citep{Hut1981,Belczynski2008}.   Meanwhile, the circularization timescale is orders of magnitude longer at $\approx$\,2~Myr, which is still only a small fraction of the secondary's pre-MS lifetime of $\approx$\,10~Myr.  Hence, it is not surprising that all of our eclipsing binaries with $P$ $<$ 4 days have been circularized, while three systems with $P$~$>$~4~days are in slightly eccentric orbits with $e$~$\approx$~0.03\,-\,0.06.

B-type MS stars have radiative envelopes, and so tidal damping is not as efficient.  Although the B-type MS primaries may spin independently from the orbital periods, we assume for simplicity that they are also tidally locked with the orbit (see also discussion of B-type MS slow rotators in \S3.1). B-type MS stars become oblate only if they rotate close to their break-up speed or nearly fill their Roche lobes \citep{Ekstrom2008}.  Fortunately, young B-type MS stars typically rotate more slowly than their break-up speed \citep{Abt2002,Ekstrom2008,Levato2013}, and the B-type MS primaries in our eclipsing binaries are well-detached from their Roche lobes.  Even if the B-type MS primaries are not already synchronized with the orbit, their true shapes will differ only slightly from our model assumptions.  For example, a B-type MS primary that quickly spins at $v_{\rm surface}$~=~300~km~s$^{-1}$ will have an equatorial radius that is only 6\% larger than its polar radius \citep{Ekstrom2008}.

\begin{figure}[t!]
\centerline{
\includegraphics[trim = 1.2cm 0.2cm 1.2cm 0.2cm, clip=true, width=3.4in]{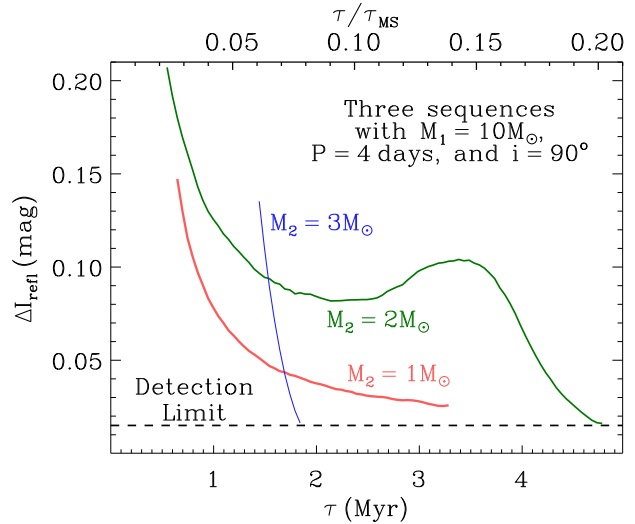}}
\caption{Reflection effect amplitude $\Delta I_{\rm refl}$ as a function of age~$\tau$ for three secondary masses $M_2$.  Above is time in units of the MS lifetime $\tau_{\rm MS}$ = 24\,Myr of the {\it primary} B-type MS star.  We show only the portions of the evolution where the light curve properties satisfy our selection criteria.  At early times $\tau$~$\lesssim$~(0.02\,$\mbox{-}$\,0.06)\,$\tau_{\rm MS}$, the companions have Roche-lobe fill-factors $RLFF_2$ $\gtrsim$ 80\% and are difficult to distinguish from large, evolved subgiants.  At later times $\tau$ $\gtrsim$ (0.1\,-\,0.2)\,$\tau_{\rm MS}$, the secondary becomes substantially smaller as it approaches its own MS phase of evolution.  Not only do the reflection effects fall below the detection limit of $\Delta I_{\rm refl}$~=~0.015~mag~(dashed~line), but the eclipse depths can also diminish below the sensitivity of the OGLE-III LMC observations. Extreme mass-ratio binaries $q$ $\lesssim$ 0.15 (red) produce observable eclipses only when the companions are on the early pre-MS phase of evolution.  Binaries at moderate mass ratios $q$ $\gtrsim$ 0.3 (blue) spend only $\lesssim$2\% of the primary's evolution in such a MS + pre-MS combination.  The nonmonotonic behavior in $\Delta I_{\rm refl}$ for the $M_2$~=~2\Msun\ sequence (green) is due to the complex pre-MS evolution of stars with $M$ $>$ 1.4\Msun, which undergo different processes of nuclear fusion and energy transport.  } 
\end{figure}

We show in Fig.~4  the reflection effect amplitude $\Delta I_{\rm refl}$ for three secondary masses $M_2$ = 1, 2, and 3 \Msun\ based on our \textsc{Nightfall} models and adopted evolutionary tracks.  For these sequences, we fix the other parameters at representative values of $M_1$ = 10\,\Msun, $P$~=~4~days, $i$~=~90$^{\rm o}$, and $A_2$ = 0.7.  The observable pre-MS duration of the 3\,\Msun\ companion is only $\sim$2\% the MS lifetime of the primary.  Hence, the majority of our eclipsing binaries that display reflection effects must have $q$~$<$~0.3 because the likelihood of observing a pre-MS + MS binary at larger $q$ is very low.  The radii of MS companions with $q$~$<$~0.15 are too small to produce detectable eclipses given the cadence and sensitivity of the OGLE-III observations.  We can therefore observe extreme mass-ratio eclipsing binaries only when the companion is large and still contracting on the pre-MS.

\begin{figure*}[t]
\centerline{
\includegraphics[trim = 3.4cm 0.4cm 4.5cm 0cm, clip=true, width=7.0in]{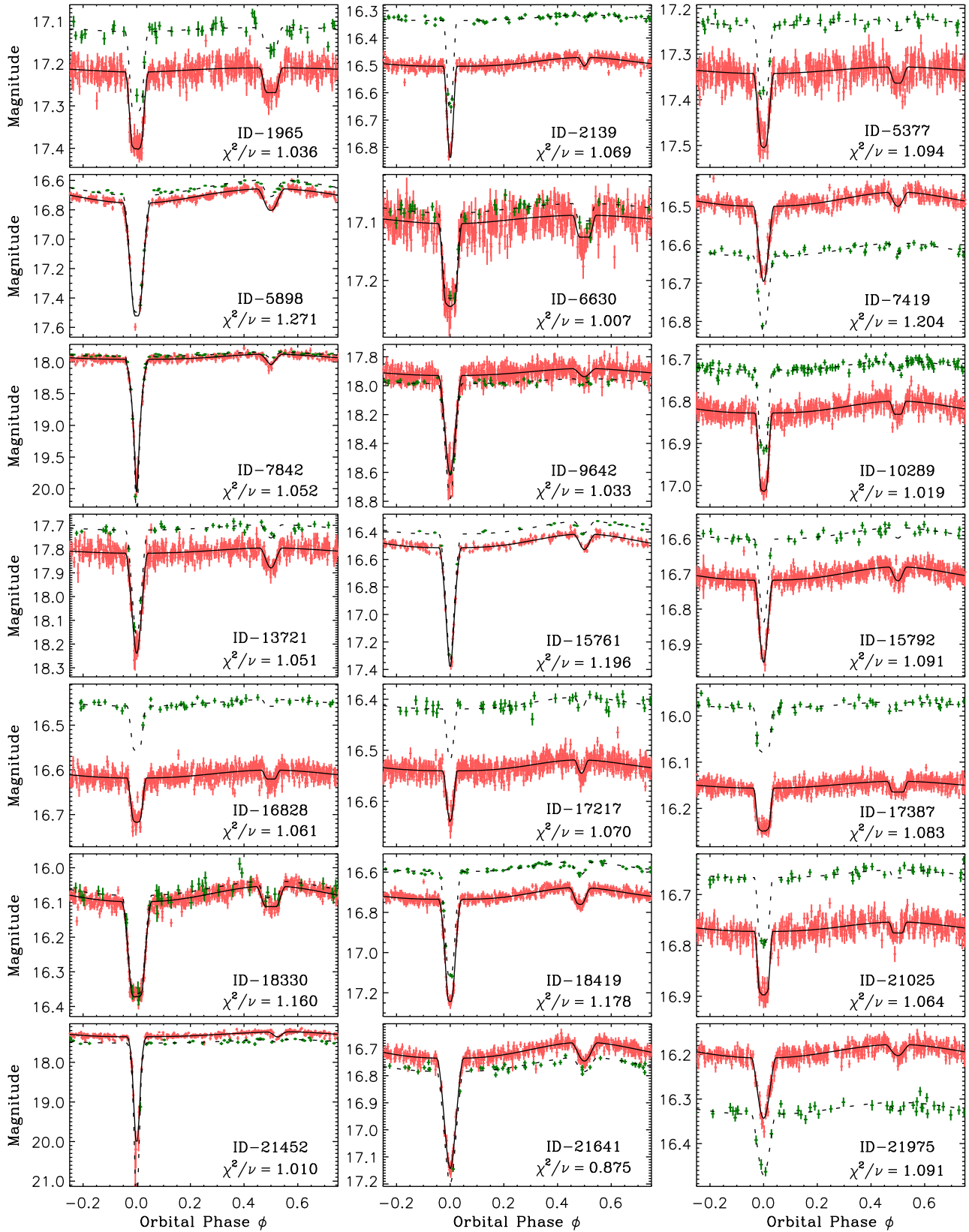}}
\caption{As shown in Fig.~1 for the prototype ID-1803, we compare the physical model fits to the  the observed light curves for the remaining 21 eclipsing binaries with B-type MS primaries and irradiated pre-MS companions. We present the physical fit parameters and statistics for all 22 systems in Tables 2-3.}
\end{figure*}

\begin{figure*}[t]\scriptsize
\begin{flushleft}
{\small {\bf Table 2}.  Best-fit model parameters and statistics for 22 eclipsing binaries with reflection effects.  The uncertainties reported in parenthesis, which include systematic uncertainties, are not necessarily symmetric around the best-fit model values.}
\end{flushleft}
\renewcommand{\tabcolsep}{1.95pt}
\vspace{-0.3cm}
\begin{tabular}{|r|l|l|r|r|l|c|r|c|c|c|c|r|c|c|c|r|}
\hline
\multicolumn{1}{|c|}{~} & \multicolumn{8}{c|}{Physical Model Properties} & \multicolumn{8}{c|}{Fit Statistics} \\
\hline
     ID~~  & ~~~$P$\,(days) & $t_{\rm o}$\,(JD-2450000) & $M_1$\,(\Msun) & $M_2$\,(\Msun) & ~$\tau$\,(Myr) & $i$\,($^{\rm o}$) & $A_2$\,(\%)\, & $A_I$\,(mag)  & ${\cal N}_I$ & ${\cal N}_{{\rm c},I}$ & $f_{\sigma,I}$ & ${\cal N}_V$ & ${\cal N}_{{\rm c},V}$ & $f_{\sigma,V}$ & $\chi^2/\nu$  &  $PTE$~ \\
 \hline
  1803~          & 3.970085~\,(7)  & ~3572.1174~\,(9)  & 10.4\,(1.6)  & 0.8\,(0.2)~  & ~\,0.7\,(0.4) & 83.6\,(1.1) & 97\,(17) & 0.32\,(4) &
                 460 & 1 & 1.17 &  41 & 0 & 1.21 & 1.062 & 0.167 \\
\hline
  1965~          & 3.175265~\,(9)  & ~3570.6285\,(17) &  7.7\,(1.2)  & 1.9\,(0.3)~  &  15\,~~(4)   & 89.8\,(1.5) & 100\,(28) & 0.22\,(3) &
                 440 & 2 & 1.09 &  45 & 0 & 1.50 & 1.036 &  0.283 \\
\hline
  2139~          & 8.462504\,(21) & ~3576.0465\,(23) & 12.7\,(2.0)  & 1.9\,(0.4)~  & ~\,0.9\,(0.5) & 84.0\,(1.3) & 29\,(15)  & 0.23\,(3) &
                 477 & 2 & 1.19 &  70 & 1 & 1.16 & 1.069 &  0.132 \\
\hline
  5377~          & 3.276373\,(10)  & ~3563.1493\,(19) & ~7.4\,(1.2)  & 1.2\,(0.2)~  & 14\,~~(5)   & 84.8\,(1.6) & 68\,(22)  & 0.19\,(3) &
                 447 & 1 & 1.11 &  43 & 0 & 1.08 & 1.094 &  0.075 \\
\hline
  5898~          & 5.323855\,(13) & ~3567.5433\,(17) & ~7.7\,(1.5)  & 3.8\,(0.8)~  & ~\,0.8\,(0.4) & 83.5\,(1.5) &  86\,(25)  & 0.07\,(3) &
                 439 & 1 & 1.30 &  44 & 2 & 1.25 & 1.271 &  $<$0.001 \\ 
\hline  
  6630~          & 3.105556~\,(9)  & ~3563.8238\,(16) &  8.3\,(1.3)  & 1.5\,(0.2)~ &   13\,~~(4)   & 86.5\,(1.3) & 100\,(27) & 0.34\,(4) &
                 410 & 1 & 1.12 &  40 & 0 & 1.05 & 1.007 &  0.450 \\
\hline
  7419~          & 4.255885~\,(9)  & ~3563.5793\,(14) & 13.7\,(2.4)  & 1.7\,(0.3)~ & ~\,5.2\,(1.5)   & 81.1\,(2.1) & 61\,(21)  & 0.66\,(7) &
                 421 & 2 & 1.12 &  40 & 0 & 1.10 & 1.204 &  0.002 \\ 
\hline
  7842~          & 3.781792~\,(8)  & ~3565.8256\,(10)  & ~6.0\,(0.9) & 2.0\,(0.3)~ & ~\,2.6\,(0.8)   & 88.5\,(0.8) & 60\,(16)  & 0.17\,(3) &
                 477 & 1 & 1.13 &  72 & 0 & 1.08 & 1.052 &  0.197 \\ 
 \hline
  9642~          & 3.913363~\,(9)  & ~3565.6120\,(14) & ~6.4\,(1.0)  & 2.3\,(0.4)~ & ~\,1.7\,(0.5)   & 81.2\,(1.4) & 29\,(19)  & 0.29\,(4) &
                 782 & 1 & 1.16 &  40 & 1 & 1.05 & 1.033 &  0.249 \\ 
\hline
 10289~          & 4.642579\,(11)  & ~3566.0280\,(15) & 11.2\,(1.8) & 0.8\,(0.2)~ & ~\,2.5\,(0.7)  & 86.9\,(0.9) & 100\,(18) & 0.32\,(4) &
                 557 & 1 & 1.07 & 115 & 0 & 1.05 & 1.019 &  0.357 \\
\hline
 13721~          & 3.122554\,(11)  & ~3558.1903\,(16) & ~6.2\,(1.0) & 1.5\,(0.3)~ &     ~\,8\,~~(2)  & 83.3\,(2.0) & 37\,(16)  & 0.14\,(3) &
                 428 & 1 & 1.05 & 40 & 0 & 1.05 & 1.051 &  0.214 \\
\hline
 15761~          & 5.310910\,(12)  & ~3581.5694\,(14) & 12.6\,(2.4) & 2.3\,(0.4)~ & ~\,1.9\,(0.5)  & 85.4\,(1.1) & 78\,(17)  & 0.30\,(4) &
                 223 & 0 & 1.74 & 21 & 0 & 1.37 & 1.196 &  0.021 \\
\hline
 15792~          & 4.317015~\,(9)  & ~3566.0986\,(15) & 11.5\,(1.8)  & 1.6\,(0.3)~ & ~\,2.8\,(1.0)   & 83.3\,(1.7) & 69\,(22)  & 0.30\,(4) &
                 600 & 1 & 1.35 & 40 & 0 & 1.41 & 1.091 &  0.056  \\
\hline
 16828~          & 3.675702\,(10)  & ~3572.0106\,(17) & 10.8\,(1.7) & 1.0\,(0.2)~ &  ~\,7\,~~(2)   & 83.7\,(2.0) & 100\,(23) & 0.23\,(3) &
                 606 & 3 & 1.18 & 47 & 0 & 1.05 & 1.061 &  0.137 \\ 
\hline
 17217$^{\rm a}$ & 5.354787\,(20) & ~3576.8028\,(26) & 11.3\,(1.8) &  1.7\,(0.4)~ &   ~\,7\,~~(3)   & 81.6\,(2.1) & 100\,(31) & 0.28\,(5) &
                 592 & 3 & 1.18 & 47 & 0 & 1.37 & 1.070 &  0.109 \\
\hline
 17387~          & 4.772901\,(14) & ~3567.4122\,(24) & 12.4\,(2.0) &  1.2\,(0.2)~ & ~\,8\,~~(2)   & 89.6\,(1.5) & 100\,(22) & 0.23\,(4) &
                 605 & 1 & 1.17 & 47 & 1 & 1.34 & 1.083 &  0.071 \\
\hline
 18330~          & 3.252921~\,(8)  & ~3561.4274\,(17) & 14.6\,(2.5) &  1.6\,(0.3)~ & ~\,5.6\,(1.5)   & 89.5\,(1.6) & 100\,(19) & 0.46\,(6) &
                 599 & 2 & 1.48 & 46 & 0 & 3.04 & 1.160 &  0.003 \\ 
\hline
 18419$^{\rm b}$ & 4.118151~\,(8)  & ~3564.3203\,(12) & 11.6\,(2.0) &  1.5\,(0.3)~ & ~\,1.6\,(0.5)   & 88.0\,(1.1) &  78\,(17) & 0.27\,(4) &
                 599 & 1 & 1.28 & 70 & 0 & 1.18 & 1.178 &  0.001 \\
\hline
 21025~          & 4.543312\,(13) & ~3581.2269\,(22) & 10.7\,(1.7) &  1.0\,(0.3)~ & ~\,6\,~~(2)   & 88.5\,(1.6) & 100\,(26) & 0.31\,(4) &
                 435 & 1 & 1.05 & 45 & 0 & 1.05 & 1.064 &  0.165 \\
\hline
 21452$^{\rm c}$ & 8.178961\,(19) & ~3541.8978\,(22) & 10.1\,(1.6) &  2.4\,(0.7)~ & ~\,0.6\,(0.4) & 89.3\,(2.2) &  67\,(17) & 0.61\,(7) &
                 337 & 1 & 1.43 & 40 & 0 & 1.88 & 1.010 &  0.436 \\
\hline
 21641~          & 3.092426~\,(7)  & ~3572.5059\,(18) & 12.7\,(2.0) &  1.9\,(0.3)~ & ~\,2.6\,(1.0)   & 82.8\,(1.6) & 51\,(19)  & 0.54\,(6) &
                 436 & 1 & 1.05 & 45 & 0 & 1.05 & 0.875 &  0.976\\
\hline
 21975~          & 3.021141~\,(9)  & ~3571.6002\,(19) & 16.0\,(2.6) &  1.7\,(0.3)~ & ~\,4.0\,(1.4)   & 77.9\,(2.3) & 36\,(21)  & 0.68\,(8) &
                 437 & 1 & 1.16 & 42 & 0 & 1.72 & 1.091 &  0.084 \\
\hline
\end{tabular}
(a): modeled with $e$ = 0.03 and $\omega$ = 230$^{\rm o}$; (b): $e$ = 0.04 and $\omega$ = 230$^{\rm o}$; (c): $e$ = 0.06 and $\omega$ =~50$^{\rm o}$; the other 19 systems have circular orbits.
\end{figure*}

\begin{figure*}[t]\footnotesize
\begin{flushleft}
{\small {\bf Table 3}. Dependent physical properties derived by using main model parameters in Table 2 in combination with Kepler's laws and stellar evolutionary tracks. The representative uncertainties are 15\% or 0.03, which ever is larger, in the mass ratios $q$ and relative ages $\tau$/$\tau_{\rm MS}$, 10\% in orbital separation $a$, radii $R_1$ and $R_2$, and Roche-lobe fill factors $RLFF_1$ and $RLFF_2$, 8\% in temperatures $T_1$ and $T_2$, 40\% in luminosities $L_1$ and $L_2$, and 0.1 mag in absolute magnitudes $M_I$ and $M_V$.}
\end{flushleft}
\renewcommand{\tabcolsep}{5.5pt}
\vspace{-0.3cm}
\begin{tabular}{|r|c|c|c|c|c|c|c|c|r|r|r|r|r|r|}
\hline
     ID~\, &   $q$ & $\tau$/$\tau_{\rm MS}$ & $a$\,(\Rsun) & $R_1$\,(\Rsun) & $R_2$\,(\Rsun) & 
     $RLFF_1$ & $RLFF_2$ & $T_1$\,(K) & $T_2$\,(K) & $L_1$\,(\Lsun) & $L_2$\,(\Lsun) & $M_I$~ & $M_V$~ \\
 \hline
  1803 & ~0.07~ & 0.03 & 23 & 3.6 & 3.1 & 0.27 & 0.77 & ~27,000 &  4,400 &  6,000~ &   3~~~~ & $-$1.7 & $-$2.1  \\
\hline
  1965 & 0.24 & 0.35 & 19 & 3.6 & 1.4 & 0.39 & 0.29 & ~23,000 &  9,900 &  3,000~ &  17~~~~ & $-$1.5 & $-$1.7  \\
\hline
  2139 & 0.15 & 0.05 & 43 & 4.2 & 3.7 & 0.19 & 0.39 & ~29,000 &  5,200 & 12,000~ &   9~~~~ & $-$2.2 & $-$2.5  \\
\hline
  5377 & 0.16 & 0.33 & 19 & 3.5 & 1.2 & 0.36 & 0.30 & ~22,000 &  6,100 &  3,000~ &   2.2~\, & $-$1.3 & $-$1.6  \\
\hline
  5898 & 0.49 & 0.02 & 29 & 3.0 & 6.3 & 0.25 & 0.73 & ~24,000 &  8,400 &  2,500~ & 180~~~~ & $-$1.8 & $-$1.9  \\
\hline  
  6630 & 0.18 & 0.41 & 19 & 3.9 & 1.3 & 0.41 & 0.31 & ~23,000 &  7,800 &  4,000~ &   6~~~~ & $-$1.7 & $-$2.0  \\
\hline
  7419 & 0.12 & 0.35 & 28 & 5.1 & 2.6 & 0.35 & 0.47 & ~30,000 &  6,200 & 18,000~ &   9~~~~ & $-$2.7 & $-$3.0  \\
\hline
  7842 & 0.33 & 0.04 & 20 & 2.7 & 2.9 & 0.29 & 0.53 & ~21,000 &  5,700 &  1,200~ &   8~~~~ & $-$0.7 & $-$0.9 \\
\hline
  9642 & 0.36 & 0.03 & 21 & 2.7 & 3.7 & 0.29 & 0.63 & ~21,000 &  5,800 &  1,400~ &  14~~~~ & $-$0.9 & $-$1.0 \\
\hline
 10289 & 0.07 & 0.13 & 27 & 4.0 & 1.5 & 0.26 & 0.34 & ~28,000 &  4,300 &  9,000~ & 0.7~\, & $-$2.0 & $-$2.3 \\
\hline
 13721 & 0.25 & 0.13 & 18 & 2.8 & 2.0 & 0.34 & 0.46 & ~20,000 &  7,200 &  1,300~ &  10~~~~ & $-$0.8 & $-$1.0  \\
\hline
 15761 & 0.18 & 0.12 & 32 & 4.3 & 4.0 & 0.27 & 0.56 & ~29,000 &  6,100 & 12,000~ &  20~~~~ & $-$2.3 & $-$2.6 \\
\hline
 15792 & 0.14 & 0.15 & 26 & 4.1 & 2.1 & 0.30 & 0.38 & ~28,000 &  5,200 &  9,000~ &   3~~~~ & $-$2.1 & $-$2.4  \\
\hline
 16828 & 0.09 & 0.33 & 23 & 4.4 & 1.2 & 0.34 & 0.30 & ~27,000 &  4,500 &  9,000~ &  0.6~\, & $-$2.1 & $-$2.4  \\
\hline
 17217 & 0.15 & 0.34 & 30 & 4.5 & 2.1 & 0.29 & 0.33 & ~27,000 & 7,800 & 10,000~ & 15~~~~ & $-$2.2 & $-$2.5 \\
\hline
 17387 & 0.09 & 0.48 & 28 & 5.2 & 1.4 & 0.33 & 0.26 & ~28,000 & 5,300 & 15,000~ &  1.4~\, & $-$2.6 & $-$2.9  \\
\hline
 18330 & 0.11 & 0.42 & 23 & 5.5 & 2.5 & 0.43 & 0.54 & ~30,000 & 6,200 & 23,000~ &  8~~~~ & $-$2.9 & $-$3.2  \\
\hline
 18419 & 0.13 & 0.08 & 25 & 4.0 & 2.4 & 0.31 & 0.48 & ~28,000 & 5,000 &  9,000~ &  3~~~~ & $-$2.0 & $-$2.4  \\
\hline
 21025 & 0.09 & 0.28 & 26 & 4.2 & 1.3 & 0.29 & 0.27 & ~27,000 & 4,600 &  8,000~ & 0.7~\, & $-$2.0 & $-$2.4  \\
\hline
 21452 & 0.24 & 0.02 & 40 & 3.6 & 5.2 & 0.20 & 0.58 & ~27,000 & 5,200 &  6,000~ & 17~~~~ & $-$1.8 & $-$2.0 \\
\hline
 21641 & 0.15 & 0.16 & 22 & 4.4 & 2.7 & 0.39 & 0.58 & ~29,000 & 5,600 & 13,000~ &  6~~~~ & $-$2.3 & $-$2.6  \\
\hline
 21975 & 0.11 & 0.34 & 23 & 5.5 & 2.5 & 0.44 & 0.56 & ~32,000 & 5,800 & 28,000~ &  6~~~~ & $-$3.0 & $-$3.3  \\
\hline
\end{tabular}
\end{figure*}

  The correction factor $f_{\sigma}(I)$ for the photometric errors we calculated in \S2.1 can differ between systems, even if they have the same magnitude.  We therefore do not use the simple relation in Eq. 5 in our physical models.  Instead, we calculate the correction factors between the catalog photometric errors and intrinsic rms scatter for each of our 22 eclipsing binaries individually.  To achieve this, we separately fit 3$^{\rm rd}$ degree polynomials across the out-of-eclipse intervals 0.05 $<$ $\phi$ $<$ 0.45 and 0.55~$<$~$\phi$~$<$~0.95 for each of the $I$-band and $V$-band light curves.  We remove all residuals that exceed 4$\sigma$, measure the rms dispersions of the remaining residuals, and then calculate the correction factors $f_{\sigma,I}$ and $f_{\sigma,V}$ between the catalog photometric errors and the measured rms scatter.  For some light curves, there are too few data points to accurately measure the correction factors, so we impose a minimum value of 1.05 for $f_{\sigma,I}$ and $f_{\sigma,V}$. For each of our 22 eclipsing binaries, we multiply the catalog photometric errors by their respective correction factors $f_{\sigma,I}$ and $f_{\sigma,V}$ when we fit our physical models.

To constrain the eight parameters in our physical models, we fit \textsc{Nightfall} synthetic light curves to the $I$-band and $V$-band data simultaneously.  As in \S2, we use a Levenberg-Marquardt method to minimize the $\chi^2$ statistic between the light curves and physical models.   We clip up to ${\cal N}_{{\rm c},I} + {\cal N}_{{\rm c},V}$ $\le$ 3 data points that deviate more than 4$\sigma$ from our best-fit model.   Since we fit both the $I$-band and $V$-band together, there are $\nu$~=~${\cal N}_I$\,$+$\,${\cal N}_V$\,$-$\,${\cal N}_{{\rm c},I}$\,$-$\,${\cal N}_{{\rm c},V}$\,$-$\,8 degrees of freedom. We compare the observed light curves to our best physical model fits in Fig.~1 (for our prototype ID-1803) and Fig.~5 (for the remaining 21 reflecting eclipsing binaries).  We present the fit parameters and statistics in Table 2, and other physical properties in Table 3.  

\subsection{Results}

For 21 of our 22 eclipsing binaries, our models have good fit statistics $\chi^2/\nu$~=~0.87\,-\,1.20.  The one remaining eclipsing binary, ID-5898, has a poor fit with $\chi^2/\nu$~=~1.27, i.e. a probability to exceed $\chi^2$ of $PTE$~$<$~0.001, most likely caused by third light contamination.  We discuss third light contamination and other systematic uncertainties in \S3.4.  

For the 21 eclipsing binaries with good fit statistics, we measure primary masses $M_1$~=~6\,-\,16\,\Msun\ appropriate for early B-type MS stars, low-mass secondaries $M_2$~=~0.8\,-\,2.4\,\Msun\ ($q$~=~0.07\,-\,0.36), young ages $\tau$~=~0.6\,-\,15~Myr, nearly edge-on inclinations $i$~=~78$^{\rm o}$\,-\,90$^{\rm o}$, secondary albedos $A_2$~=~(30\,-\,100)\%, and moderate to large dust extinctions $A_I$~=~0.14\,-\,0.68~mag. The B-type MS primaries have relative ages $\tau$/$\tau_{\rm MS}$ that span from 2\% up to 50\% their MS lifetimes.  The fits confirm these eclipsing binaries with narrow eclipses are in detached configurations with Roche-lobe fill-factors $RLFF_1$~=~0.2\,-\,0.4 and $RLFF_2$~=~0.3\,-\,0.8.  Given the orbital separations $a$~=~20\,-\,40\,\Rsun, these fill-factors correspond to physical radii $R_1$~=~2.7\,-\,5.5\,\Rsun\ and $R_2$~=~1.2\,-\,5.2\,\Rsun.  Finally, as expected for eclipsing binaries that exhibit substantial reflection effects, we find comparable radii $R_2$/$R_1$~$=$~0.3\,-\,1.4 but extreme contrasts in temperature $T_2$/$T_1$~=~0.15\,-\,0.43 and luminosity $L_2$/$L_1$~$\approx$~10$^{-4}$\,-\,10$^{-2}$.  

  Considering B-type MS stars span a narrow range of temperatures $T_1$ and radii $R_1$, the temperatures $T_2$ and radii $R_2$ of the companions are more accurately and robustly measured than their masses $M_2$ or ages $\tau$.  In Fig.~6, we compare the locations of the pre-MS companions on a Hertzsprung-Russell diagram to the theoretical Pisa evolutionary tracks \citep{Tognelli2011}.  ID-5898 is biased toward larger $L_2$ most likely due to third light contamination (see below).  ID-1965, ID-5377, and ID-6630 have small reflection effect amplitudes $\Delta I_{\rm refl}$~=~0.017\,-\,0.20~mag just above our detection limit of 0.015~mag (see Fig.~5 and Table 1).  These three systems also have shallow, flat-bottomed eclipses $\Delta I_1$~$\approx$~0.2~mag that dictate full non-grazing eclipse trajectories and ratio of radii $R_2$/$R_1$ $\approx$ 0.3\,-\,0.4.  The companions in these three eclipsing binaries are therefore small, warm, late pre-MS or zero-age MS stars with relatively older ages $\tau$~$\approx$~13\,-\,15~Myr (Table~2).  Nonetheless, these three systems still have small secondary masses $M_2$~=~1.2\,-\,1.9\,\Msun\ ($q$~=~0.16\,-\,0.24), so we keep these eclipsing binaries in our statistical sample.  

The remaining 18 systems have deeper eclipses and/or larger reflection effect amplitudes, which dictate the companions are larger and/or cooler.  The secondaries in these 18 eclipsing binaries are inconsistent with zero-age MS stars (Fig.~6), but instead must be primordial pre-MS stars with young ages $\tau$ $=$ 0.6\,-\,8~Myr and small masses $M_2$ = 0.8\,-\,2.4\,\Msun.  The majority of these companions have developed a radiative core and are evolving with nearly constant $R_2$ on the Henyey track \citep{Siess2000,Tognelli2011}. A few secondaries are still fully convective and contracting on the Hyashi phase of the pre-MS.  According to our adopted pre-MS evolutionary tracks \citep{Tognelli2011}, eleven of our pre-MS secondaries have not yet initiated stable nuclear burning in their cores but are powered completely by gravitational energy.

\begin{figure}[t!]
\centerline{
\includegraphics[trim = 0.4cm 0.2cm 0.9cm 0.2cm, clip=true, width=3.7in]{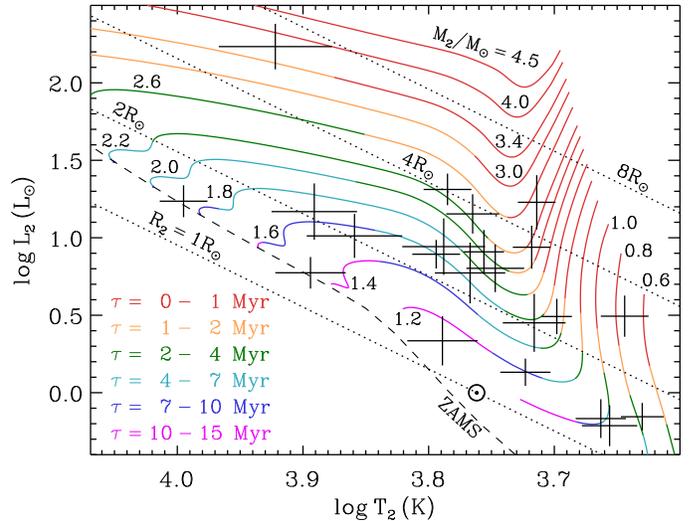}}
\caption{Hertzsprung-Russell diagram of the companions in our 22 eclipsing binaries.  We compare the dependent model properties $T_2$ and $L_2$ to the Pisa pre-MS tracks \citep{Tognelli2011} used to constrain the parameters of the observed systems.  We display evolutionary tracks for secondary masses $M_2$ = 0.6\,-\,4.5\,\Msun, where the colors indicate the ages of the pre-MS stars.  We also show lines of constant radius (dotted) and the zero-age MS (dashed).  ID-5898, which has the worst model fit statistic $\chi^2/\nu$ = 1.27 and is probably contaminated by a third light source, is toward the top left.  The three systems to the bottom left (ID-1965, ID-5377, and ID-6630) have small reflection effect amplitudes $\Delta I_{\rm refl}$ = 0.017\,-\,0.20~mag, shallow eclipses $\Delta I_1$ = 0.2 mag, and companions that are consistent with the zero-age MS.  The remaining 18 eclipsing binaries have companions that are larger and/or cooler and therefore definitively pre-MS stars.  The observed systems cluster on the Henyey track near $T_2$ $\approx$ 6,000\,K and $L_2$ $\approx$ 10\,\Lsun, which corresponds well to where large pre-MS stars with $R_2$ $\approx$ 2\,-\,4\,\Rsun\ are longest lived and therefore have the highest probability of producing detectable reflection effects.}
\end{figure}

\subsection{Systematic Uncertainties}

The one system with a poor model fit, i.e. ID-5898,  converges toward a solution with a high-mass secondary $M_2$ = 3.8\,\Msun\ ($q$~$=$~0.49), young age $\tau$ $\approx$ 0.8 Myr, and small dust extinction $A_I$ = 0.07.  We find four reasons to suspect this system suffers from contamination with a third light source, most likely a hot late-B/early-A tertiary companion.  First, the amplitude of the reflection effect in ID-5898 appears to be color dependent with $\Delta I_{\rm refl}$ = 0.10 mag and $\Delta V_{\rm refl}$~=~0.07~mag~(see~Fig.~5).  The decrease in  $\Delta V_{\rm refl}$ is most likely caused by stellar blending with a third light source that is relatively hot and brighter in the V-band.  Second, the measured dust extinction $A_I$~=~0.07 is smaller than that compared to dust reddening estimates of young stars along similar lines-of-sight \citep[][see also below]{Zaritsky2004}.  Third light contamination from a hot source would artificially shift the observed color toward the blue and bias our dust reddening measurement toward smaller values.  Third, extra light would diminish the primary eclipse depth $\Delta I_1$. This would mainly lead to an underestimation of the inclination $i$, but may also cause us to overestimate $L_2$, $R_2$, and $M_2$.  Considering the other 21 companions have $L_2$ $\lesssim$ 20\,\Lsun, $R_2$ $\lesssim$ 5.2\,\Rsun, and $M_2$ $\lesssim$ 2.4\,\Msun, the measurements of $L_2$ $\approx$ 180\,\Lsun, $R_2$ $\approx$ 6.3\,\Rsun, and $M_2$~$\approx$~3.8\,\Msun\ for ID-5898 are clear outliers and indicative of third light contamination.  Finally, because third light contamination can bias our light curve solution to larger $L_2$ and $R_2$, our measured $\tau$ is also shifted toward younger ages. Of the four eclipsing binaries in our sample with age estimates $\tau$ $\lesssim$ 1 Myr, only ID-5898 is not embedded in a bright and/or compact H\,\textsc{ii} region (see \S4). 

Considering the above, we remove ID-5898 when discussing correlations (\S4) and the intrinsic binary fraction (\S5).  Nonetheless, ID-5898 is phenomenologically similar to the other 21 eclipsing binaries in our sample, and it most likely contains a low-mass pre-MS companion.  We therefore still include this system in our total sample of 22 reflecting eclipsing binaries.  We are simply unable to accurately constrain the physical properties of this system because of systematic effects most likely caused by third light contamination.  Even if ID-5898 has a true mass ratio $q$ $<$ 0.25, the addition of this one object to the 19 measured systems with $q$ = 0.07\,-\,0.25 would have a negligible effect on our statistics.

For each of our 22 eclipsing binaries, we calculate the covariance matrix and measurement uncertainties in our eight physical model parameters.  However, most of our measured physical properties are dominated by systematic errors.  In the following, we quantify the magnitudes and directions of various sources of systematic errors:

\begin{enumerate}[leftmargin=*]

\item Bolometric corrections.  For our hot B-type MS primaries, the bolometric corrections are large and typically uncertain by 0.2\,-\,0.3 mag \citep{Pecaut2013}.  This dictates the primary luminosities $L_1$ are uncertain by at least $\approx$20\,-\,30\%, and therefore the inferred primary masses $M_1$ have systematic uncertainties of at least $\approx$10\%.  However, if we were to systematically overestimate or underestimate $M_1$, we would also bias our inferred $M_2$ in the same direction.  This is because the measured ratio of eclipse depths $\Delta I_2$/$\Delta I_1$ mainly determines the luminosity contrast $L_2$/$L_1$ and therefore the mass ratio $q = M_2/M_1$.  Hence, our measured mass ratios $q$ are relatively insensitive to the uncertainties in the bolometric corrections.  

\item  Color indices.   Given a surface temperature $T_2$,  the intrinsic colors $(V-I)_{\rm o}$ of hot B-type MS stars are uncertain by $\approx$0.02 mag \citep{Pecaut2013}.  The zero-point calibrations in the measured OGLE-III LMC colors are also uncertainty by $\approx$0.01\,-\,0.02 mag \citep{Udalski2008}.  Our measured dust extinctions $A_I$ therefore have a minimum systematic error of $\approx$0.03 mag.

\item Dust reddening law.  The coefficient in our adopted dust reddening law $E(V-I)$ = 0.7$A_I$ has a systematic error of $\sim$10\% \citep{Cardelli1989, Fitzpatrick1999,Ngeow2005}.  The inferred dust extinctions $A_I$ are also uncertain by this factor.

\item Evolutionary tracks.  Given a luminosity $L_1$ of the primary B-type MS star, the primary masses are uncertain by $\approx$10\% according to the stellar evolutionary tracks \citep{Dotter2008,Bertelli2009}.  Rotating models of young B-type MS stars are only 4\% cooler than their non-rotating counterparts \citep{Ekstrom2008b}.  This implies a 5\% systematic uncertainty in the masses of the B-type MS stars due to the uncertainty in the rotation rates. For the pre-MS companions in our eclipsing binaries, we compare pre-MS models based on four different calculations \citep{Siess2000,Dotter2008,diCriscienzo2009,Tognelli2011}. For ages $\tau$ $\gtrsim$ 1 Myr and masses $M_2$ $>$ 1.3, all pre-MS evolutionary tracks agree fairly well with typical errors of $\approx$15\% in mass and $\approx$25\% in age. At younger ages and lower masses, the systematic uncertainties increase to $\delta \tau$ $\approx$ 0.3 Myr and $\delta M_2$ $\approx$ 0.2\Msun.

\item Irradiation effects. The luminosity received by the pre-MS companion from the B-type MS star is comparable to the intrinsic luminosity of the pre-MS star itself.  This may cause the companion to enlarge, especially if it has a convective envelope and the albedo is measurably less than unity.  This effect has been studied in the context of low-mass X-ray binaries in which a hot accretion disk around a compact object irradiates a cool, low-mass donor \citep{Podsiadlowski1991,Ritter2000}.  If the irradiation effects are on one side, as they are in X-ray binaries as well as in our eclipsing binaries, then the radius of the companion increases by only $\approx$5\% \citep{Ritter2000}.  Instead of becoming stored in the interior of the star, the intercepted energy quickly diffuses laterally to the unirradiated side and subsequently lost via radiation.  This $\approx$5\% systematic effect in radius is smaller than the uncertainties due to the evolutionary tracks discussed above.  Most importantly, irradiation effects would shift the pre-MS companions toward larger radii and luminosities, so that we would have  overestimated, not underestimated, their masses.  Our conclusion that the companions in our eclipsing binaries are low-mass pre-MS stars is therefore not affected by irradiation effects.  

\item Zero-point age calibration. Although our eclipsing binaries are detached from their Roche lobes and are currently evolving relatively independently from each other, they most likely experienced prior coevolution.  In particular, the two components probably competed for accretion in the same circumbinary disk \citep{Bate2002}. Isolated T Tauri pre-MS stars with masses $\approx$\,1\,-\,3\,\Msun\ still have thick circumstellar disks at ages $\tau$~$\approx$~0.5\,-\,5~Myr \citep{Hartman2009}.  There is no evidence for circumstellar disks in the photometric light curves of our 22 eclipsing binaries in the LMC as we observe in nearby BM Orionis \citep[][see~\S2]{Windemuth2013}.    The absence of circumstellar disks in our eclipsing binaries demonstrates that the pre-MS companions formed differently than they would have in isolation.  Nonetheless, most of the mass of a solar-type star is accreted at very early stages $\tau$~$\lesssim$~0.2~Myr \citep{Hartman2009}.   Moreover, the theoretical evolutionary tracks \citep{Siess2000,Dotter2008,diCriscienzo2009,Tognelli2011}  assume pre-MS stars evolve with constant mass, which better describe our low-mass pre-MS companions without disks than isolated low-mass pre-MS stars with disks.  

The time of initial pre-MS contraction and observability, sometimes called the birthline \citep{Palla1990}, can differ by 0.2 Myr between components in the same binary system \citep{Stassun2008}.  Fortunately, the initial contraction phases are extremely rapid, and so the zero-point age calibration is uncertain by at most $\approx$0.4~Myr (see also fourth item in this list).  Finally, we measured the ages $\tau$ and masses $M_2$ of the {\it companions} according to their properties $T_2$ and $R_2$ (Fig.~6).  We inferred these companion properties from $T_1$, $R_1$, and the light curve characteristics.  Because $T_1$ and $R_1$ of the B-type MS primaries evolve much more slowly than $T_2$ and $R_2$ of the low-mass pre-MS companions, then our models are not too sensitive to the age of the primary.  Even if the primary was slightly older or younger than the companion, we would still measure the same age $\tau$ and mass $M_2$ for the secondary.  In short, the current properties of our pre-MS companions with ages $\tau$~$>$~0.6 Myr are primarily dictated by their masses, with little dependence on the presence of a disk, prior coevolution at $\tau$ $\lesssim$ 0.4 Myr, or age of the primary.

\item  Eclipsing binary models.  For the same physical parameters, we compare our best-fit models produced by \textsc{Nightfall} with light curves generated by the eclipsing binary software \textsc{Phoebe} \citep{Prsa2005}.  We find only slight differences, typically caused by the different treatment of limb-darkening and albedo between the two packages.

\item Third light contamination. Our measured physical properties can deviate beyond the calculated uncertainties if the photometric light curves include a third light source that is brighter than $\gtrsim$10\% the luminosity of the B-type MS primary.  In \citet{Moe2013}, we measured the spatial density of bright stars, typically giants, in the LMC.  We determined the probability that a luminous B-type MS eclipsing binary is blended with such a bright foreground or background star is only $\approx$\,5\%.  Most close binaries are orbited by an outer tertiary component \citep{Tokovinin2006}, but wide companions are weighted toward small mass ratios \citep{Abt1990,Shatsky2002}.  Hence, the probability that our eclipsing binaries are orbited by a bright, massive late-B/early-A tertiary component is only $\approx$10\% \citep{Moe2013}.  Given our sample of 22 reflecting eclipsing binaries, we expect only one to be blended with a background or foreground cool giant, and possibly two to contain a hot luminous tertiary companion. ID-5898 probably experiences the latter of these two types of third light contamination.  In addition, our model for ID-7419 results in a moderately poor fit statistic $\chi^2/\nu$ = 1.20, low inclination $i$ $\approx$ 81$^{\rm o}$, and large dust extinction $A_I$ $\approx$ 0.7 mag.  ID-7419 is most likely contaminated with a cool foreground or background giant.  Third light contamination causes us to overestimate, not underestimate, the secondary masses $M_2$, and typically results in larger $\chi^2/\nu$ statistics.  Most importantly, third light contamination affects only two to three individual systems in our sample, not our entire population like the previously discussed sources of systemic errors.
\end{enumerate}

Based on the above, we add in quadrature to our statistical measurement uncertainties the following systematic errors.  $M_1$: 15\% relative error;  $M_2$: 15\% relative error or absolute error of 0.2\,\Msun, whichever is larger; $\tau$: 25\% or 0.4 Myr, whichever is larger; and $A_I$: 10\% or 0.03 mag, whichever is larger. We propagate these systematic uncertainties in $M_1$, $M_2$, $\tau$, and $A_I$  into the other model parameters $P$, $t_{\rm o}$, $i$, and $A_2$ according to the covariance matrix.   In Table 2, the values in parenthesis represent the total uncertainties, including systematic errors, in the final decimal places of our eight model parameters.  Note that we list the best-fit model parameters in Table 2, and so the reported  uncertainties are not necessarily symmetric around the best-fit values.  The uncertainties in $P$ and $t_{\rm o}$ primarily derive from the observed light curves with little contribution from the systematic errors.  The uncertainties in the secondary albedos $A_2$ are quite large, but the other physical properties are relatively independent of this parameter.  Typical errors in the mass ratios $q$ and relative ages $\tau/\tau_{\rm MS}$ are 15\% or 0.03, whichever is larger.  The uncertainties in the luminosities $L_1$ and $L_2$ are $\approx$40\%, which are primarily due to the systematic uncertainties in the bolometric corrections discussed above.  Given the precisely measured orbital periods and the uncertainties in the masses and evolutionary tracks, the uncertainties in the orbital separations $a$, radii $R_1$ and $R_2$, and Roche-lobe fill-factors $RLFF_1$ and $RLFF_2$ are $\approx$10\% according to Kepler's third law.  Finally,  given the uncertainties in the radii and luminosities, the representative errors in the temperatures $T_1$ and $T_2$ are $\approx$8\% according to the Stefan-Boltzmann law.  Although the systematic errors dominate our total uncertainties,  our conclusions that 19 of our eclipsing binaries with reflection effects have small mass ratios $q$ $\lesssim$ 0.25 and young ages $\tau$~$\lesssim$~15~Myr are robust. 

As a consistency check, we compare our fitted dust reddening measurements $E(V-I)$ = 0.7$A_I$  to the \citet{Zaritsky2004} LMC dust reddening maps.  Specifically, we compile the  \citet{Zaritsky2004} $A_V$ values toward hot, young LMC stars within one arcminute of each of our 22 eclipsing binaries.  We typically find $\approx$10\,-\,25 such systems in their  database that are this close to our eclipsing binaries.  For each area, we calculate the mean extinction and standard deviation of the mean extinction.  We convert these V-band extinction values into dust reddenings $E(V-I)$ = 0.41$A_V$ using are adopted reddening law.  Finally, we add in quadrature to the measurement uncertainties a systematic error of 0.03 mag (see second item in list above).  In Fig.~7, we compare our values and uncertainties for $E(V-I)$ to those we compiled from \citet{Zaritsky2004}.  Only ID-5989 and ID-7419, which have the largest $\chi^2/\nu$ statistics most likely caused by third light contamination, have dust reddening measurement that are discrepant with the \citet{Zaritsky2004} values.  As discussed above, third light contamination will bias our dust reddening measurements toward larger or smaller values, depending on the color of the third light source.  For the remaining 20 of our eclipsing binaries, the two dust reddening estimates are in agreement, which demonstrates the reliability of our eclipsing binary models.

\begin{figure}[t!]
\centerline{
\includegraphics[trim = 0.4cm 0.2cm 1.1cm 0.2cm, clip=true, width=3.5in]{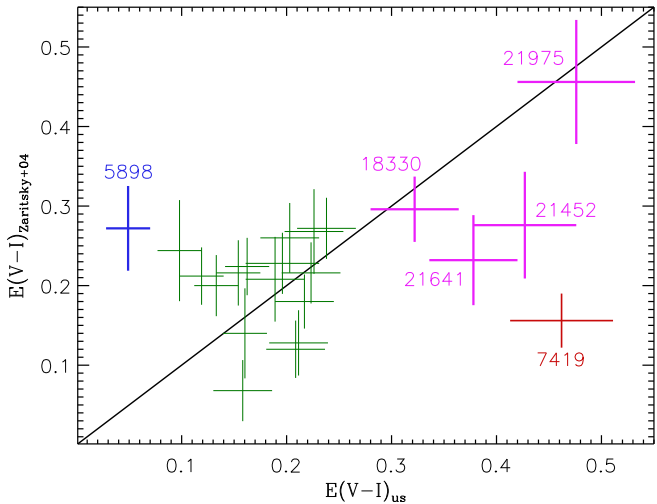}}
\caption{Comparison of our fitted dust reddening measurements to values obtained by \citet{Zaritsky2004} of hot young stars along similar lines-of-sight to our 22 nascent eclipsing binaries.  The majority of systems (green) are only slightly reddened with $E(V-I)$ $\approx$ 0.1\,-\,0.3.  The four highly reddened eclipsing binaries (magenta) with $E(V-I)$ $\approx$ 0.3\,-\,0.5 are in the eastern portions of the LMC and embedded in bright, dusty H\,\textsc{ii} regions, e.g. ID-21452 is in 30 Doradus. The two systems that deviate more than 3$\sigma$ from the \citet{Zaritsky2004} measurements, ID-5989 (blue) and ID-7419 (red), happen to be the two eclipsing binaries with the poorest model fits to the observed light curves, $\chi^2/\nu$ = 1.27 and  $\chi^2/\nu$ = 1.20, respectively.  This indicates third light contamination dominates the systematic errors in these two systems, while the other 20 eclipsing binaries are relatively free from third light contamination. } 
\end{figure}

\section{Association with H\,\textsc{ii} Regions}

Because our eclipsing binaries with reflection effects contain pre-MS companions, they should be systematically younger than their non-reflecting counterparts.  To test this prediction, we check for correlations between the coordinates of the eclipsing binaries and positions of star-forming H\,\textsc{ii}~regions in the LMC. In Table~4, we list various properties of our 22 eclipsing binaries, including their identification numbers and coordinates from the OGLE-III LMC eclipsing binary catalog \citep{Graczyk2011}.  To perform our statistical analysis below, we utilize the coordinates, sizes, and position angles of the 1,164 H\,\textsc{ii} regions in the \citet{Bica1999} catalog designated as class NA or NC, i.e. stellar associations and clusters, respectively, clearly related to emission nebulae. We report in Table 4 the properties of the H\,\textsc{ii} regions with which 20 of our reflecting eclipsing binaries are associated.  This includes the projected offset $r$ (in pc) between the eclipsing binaries and the centers of the H\,\textsc{ii} regions, the physical radii $\langle r \rangle _{\rm H\,II}$ (in pc) of the H\,\textsc{ii} regions, and the H\,\textsc{ii} region catalog identification numbers and names from \citet{Bica1999}.  We define the mean physical radius to be $\langle r \rangle _{\rm H\,II}$ =  $\sqrt{A \times B}/2$, where $A$ and $B$ are the major and minor axes provided by \citet{Bica1999} projected at the distance $d$~=~50~kpc to the LMC.

\begin{figure}[b]
\centerline{
\includegraphics[trim = 0.0cm 0.0cm 0.0cm 0.0cm, clip=true, width=3.4in]{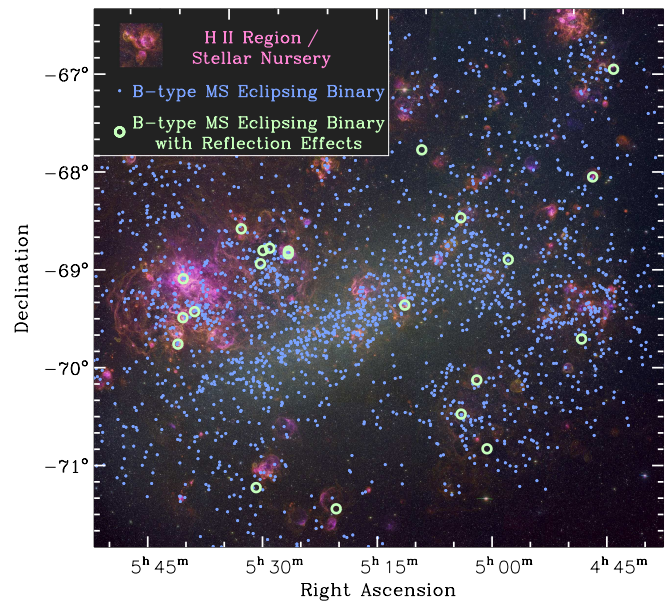}}
\caption{The positions of the 2,206 eclipsing binaries with B-type MS primaries and orbital periods $P$ = 3\,-\,15~days (blue dots) and the subset of 22 systems that exhibit pronounced reflection effects with pre-MS companions (green circles) superimposed on a narrow-band color image of the LMC taken from the Magellanic Cloud Emission Line Survey \citep{Smith2005}.  The largest concentration of normal B-type MS eclipsing binaries is in the central bar of the LMC, while those displaying reflection effects typically reside in star-forming H\,\textsc{ii} regions. Relative to their non-reflecting counterparts, the positions of our 22 reflecting eclipsing binaries are correlated with the positions of H\,\textsc{ii} regions at the 4.1$\sigma$ confidence level.  This demonstrates our 22 eclipsing binaries that exhibit reflection effects are systematically younger, which reinforces our conclusion that they contain low-mass pre-MS secondaries.}
\end{figure}

\begin{table*}[t]\footnotesize
\begin{flushleft}
{\small {\bf Table 4}. Coordinates and properties of the 22 eclipsing binaries with reflection effects, and their association with H\,\textsc{ii} regions.}
\end{flushleft}
\vspace{-0.3cm}
\renewcommand{\tabcolsep}{4.6pt}
\begin{tabular}{|r|c|c|c|c|r|r|r|r|l|}
\hline
 ~~~ID~~ & ~~RA\,(J2000)~~  &  ~~DE\,(J2000)~~ &  ~$\Delta I_1$~ &  ~$\Delta I_{\rm refl}$~ & $\tau$ (Myr) & $r$ (pc) & $\langle r \rangle_{\rm H\,II}$ (pc) & H\,\textsc{ii} ID & H\,\textsc{ii} Name   \\
\hline
 1803 & 4h\,51m\,58.23s & $-$66$^{\rm o}$\,57$'$\,00.0$''$ & 0.64  & 0.138  & 0.7~~  &   25~~ &   8~~~~~ & 351~ & NGC1714 in Shapley-VI   \\
\hline
 1965 & 4h\,52m\,34.89s & $-$69$^{\rm o}$\,42$'$\,24.7$''$ & 0.20  & 0.017  & 15~~~~\,  &  140~~ & 400~~~~~ & 470~ & SGshell-LMC7    \\
\hline
 2139 & 4h\,53m\,05.17s & $-$68$^{\rm o}$\,03$'$\,03.4$''$ & 0.33  & 0.033  & 0.9~~  &    3~~ &   6~~~~~ & 418~ & HDE268680 in NGC1736 \\
\hline
 5377 & 5h\,01m\,44.99s & $-$68$^{\rm o}$\,53$'$\,43.2$''$ & 0.17  & 0.020  & 14~~~~\,   &       &       &      &   \\
\hline
 5898 & 5h\,02m\,58.72s & $-$70$^{\rm o}$\,49$'$\,44.7$''$ & 0.83  & 0.098  & 0.8~~  &  380~~ & 400~~~~~ & 1172~ & Shapley-VIII      \\
\hline
 6630 & 5h\,04m\,39.43s & $-$70$^{\rm o}$\,07$'$\,33.3$''$ & 0.16  & 0.019  & 13~~~~\,  &   19~~ &  26~~~~~ & 1331~ & BSDL552 in LMC-DEM68   \\
\hline
 7419 & 5h\,06m\,21.44s & $-$70$^{\rm o}$\,28$'$\,27.5$''$ & 0.19  & 0.037  & 5.2~~  &  290~~ & 400~~~~~ & 1172~ & Shapley-VIII     \\
\hline
 7842 & 5h\,07m\,17.97s & $-$68$^{\rm o}$\,28$'$\,03.8$''$ & 1.73  & 0.083 & 2.6~~  &    5~~ &  16~~~~~ & 1572~ & BSDL657 in LMC-DEM76      \\
\hline
 9642 & 5h\,11m\,46.29s & $-$67$^{\rm o}$\,46$'$\,25.1$''$ & 0.70  & 0.050  & 1.7~~  &       &       &       &            \\
\hline
10289 & 5h\,13m\,23.92s & $-$69$^{\rm o}$\,21$'$\,37.3$''$ & 0.20  & 0.034  & 2.5~~ &    6~~ &  11~~~~~ & 2124~ & NGC1876 in SL320     \\
\hline
13721 & 5h\,21m\,51.38s & $-$71$^{\rm o}$\,26$'$\,31.5$''$ & 0.42  & 0.025 & 8~~~~\,  &   80~~ & 120~~~~~ & 3018~ & LMC-DEM164 in SGshell-LMC9     \\
\hline
15761 & 5h\,26m\,35.54s & $-$68$^{\rm o}$\,48$'$\,35.7$''$ & 0.85  & 0.108  & 1.9~~  &    5~~ &   3~~~~~ & 3598~ & LMC-N144B in SL476    \\
\hline
15792 & 5h\,26m\,37.37s & $-$68$^{\rm o}$\,50$'$\,09.2$''$ & 0.23  & 0.039  & 2.8~~  &   17~~ &  11~~~~~ & 3635~ & NGC1970 in SL476      \\
\hline
16828 & 5h\,28m\,40.41s & $-$68$^{\rm o}$\,46$'$\,42.8$''$ & 0.11  & 0.018  & 7~~~~\,  &  170~~ & 200~~~~~ & 3759~ & Shapley-II in SGshell-LMC3    \\
\hline
17217 & 5h\,29m\,27.69s & $-$68$^{\rm o}$\,48$'$\,09.9$''$ & 0.10  & 0.022  & 7~~~~\,  &  180~~ & 200~~~~~ & 3759~ & Shapley-II in SGshell-LMC3   \\    
\hline
17387 & 5h\,29m\,49.28s & $-$68$^{\rm o}$\,56$'$\,17.6$''$ & 0.09  & 0.017  & 8~~~~\,  &  150~~ & 200~~~~~ & 3759~ & Shapley-II in SGshell-LMC3   \\   
\hline
18330 & 5h\,31m\,44.95s & $-$68$^{\rm o}$\,34$'$\,52.5$''$ & 0.30  & 0.056  & 5.6~~  &   11~~ &  11~~~~~ & 4256~ & BSDL2159 in LMC-DEM227    \\
\hline
18419 & 5h\,31m\,55.78s & $-$71$^{\rm o}$\,13$'$\,32.0$''$ & 0.55  & 0.059  & 1.6~~  &   30~~ &   4~~~~~ & 4389~ & LMC-N206D in SGshell-LMC9    \\
\hline
21025 & 5h\,37m\,45.75s & $-$69$^{\rm o}$\,25$'$\,38.8$''$ & 0.13  & 0.025  &  6~~~~\,  &   48~~ &  56~~~~~ & 5056~ & LMC-DEM261 in LH96    \\
\hline
21452 & 5h\,38m\,43.99s & $-$69$^{\rm o}$\,05$'$\,29.6$''$ & 2.82  & 0.124  & 0.6~~  &    8~~ &  26~~~~~ & 5112~ & 30 Doradus in NGC2070      \\
\hline
21641 & 5h\,39m\,10.90s & $-$69$^{\rm o}$\,29$'$\,20.2$''$ & 0.41  & 0.062  & 2.6~~  &   12~~ &  27~~~~~ & 5140~ & NGC2074 in LMC-N158      \\
\hline
21975 & 5h\,40m\,03.86s & $-$69$^{\rm o}$\,45$'$\,32.3$''$ & 0.14  & 0.030  & 4.0~~  &    5~~ &  10~~~~~ & 5252~ & NGC2084e in LMC-N159    \\
\hline
\end{tabular}

\end{table*}

In Fig.~8, we show the coordinates of the 2,206 B-type MS eclipsing binaries in our full sample, and the 22 systems that exhibit reflection effects with large, pre-MS companions.  In the background of Fig.~8, we display an image of the LMC taken from the Magellanic Cloud Emission Line Survey \citep{Smith2005}, where the star-forming  H\,\textsc{ii} regions are clearly visible. Based on the \citet{Bica1999} catalog,  only 16\% of normal B-type MS eclipsing binaries have projected distances $r$~$<$~30\,pc from the centers of H\,\textsc{ii} regions.  In contrast, 13 of our 22 systems with reflection effects, i.e. (59\,$\pm$10\%), are situated this close to such stellar nurseries.  These values differ at the 4.1$\sigma$ significance level, demonstrating that B-type MS eclipsing binaries with reflection effects are dramatically younger. 

Similarly, only 4.8\% of B-type MS eclipsing binaries are located in centrally condensed H\,\textsc{ii} regions with mean physical radii $\langle r \rangle_{\rm H\,II}$ $=$ 3\,-\,30\,pc.  Meanwhile, 10 of the 22 systems with reflection effects, i.e. (45\,$\pm$\,11)\%, are embedded in such star-forming environments.  In addition, 10 of the remaining 12 reflecting eclipsing binaries are associated with extended, more diffuse H\,\textsc{ii} regions with $\langle r \rangle_{\rm H\,II}$ $>$ 30\,pc. These statistics demonstrate that our B-type MS eclipsing binaries with reflection effects are relatively young with ages $\tau$ $\approx$ 1\,-\,8 Myr that are comparable to the lifetimes of H\,\textsc{ii} regions.   

Only 2 of our 22 eclipsing binaries with reflection effects do not appear to be associated with an  H\,\textsc{ii} region.  One of these systems, ID-5377, is relatively old at $\tau$~=~14\,$\pm$\,5~Myr (Table~4), and so it is not unexpected that it is relatively remote from a site of  active star formation.  In contrast, the other eclipsing binary that is not in an H\,\textsc{ii} region, ID-9642, is relatively young at $\tau$ = 1.7\,$\pm$0.5~Myr.  We speculate that this eclipsing binary with a B-type MS primary may have formed in relative isolation without nearby O-type stars to ionize the surrounding gas \citep[see][]{deWit2004,Parker2007}.  As another possibility, the young ID-9642 may be embedded in a compact H\,\textsc{ii} region with $\langle r \rangle_{\rm H\,II}$ $\lesssim$ 1\,pc that is below the resolution limit of ground-based surveys and therefore not in the \citet{Bica1999} catalog.  In any case, the fact that 20 our our 22 eclipsing binaries are associated with  H\,\textsc{ii} regions reinforces our conclusion that the majority of the companions are young, low-mass, pre-MS stars.

The positions of our 22 eclipsing binaries and their associations with  H\,\textsc{ii} regions also corroborate the reliability of our eclipsing binary models.  For example, ID-15761 and ID-15792 are both associated with the same H\,\textsc{ii} region SL476.  For these two systems, we derived ages  $\tau$ = 1.9\,$\pm$\,0.5 Myr and $\tau$ = 2.8\,$\pm$\,1.0 Myr, respectively, that are consistent with each other, and dust extinctions $A_I$~=~0.30\,$\pm$\,0.04~mag that match each other.  Similarly, ID-16828, ID-17217, and ID-17387 are all in the large diffuse H\,\textsc{ii} region Shapley-II with $\langle r \rangle_{\rm H\,II}$~$\approx$~200 pc.  These three eclipsing binaries have slightly older ages $\tau$ $\approx$ 7\,-\,8 Myr and consistently smaller dust extinctions $A_I$ $\approx$ 0.23\,-\,0.28~mag.  Our youngest three eclipsing binaries with reliable age estimates $\tau$~$\lesssim$~1~Myr, i.e. ID-1803, ID-2139, and ID-21452, are all associated with centrally condensed H\,\textsc{ii} regions with $\langle r \rangle_{\rm H\,II}$ $\lesssim$ 30 pc.  Alternatively, our three oldest systems with $\tau$ $\approx$ 13\,-\,15 Myr and companions close to the zero-age MS, i.e. ID-1965, ID-5377, and ID-6630, are either not associated with star-forming environments or are in relatively large and/or diffuse H\,\textsc{ii} regions.

We now examine these correlations between eclipsing binary parameters and the properties of the  H\,\textsc{ii} regions in which they reside in a more statistical manner. In Table 4, we list the observed primary eclipse depths $\Delta I_1$ and reflection effect amplitudes $\Delta I_{\rm refl}$ from Table 1, and the modeled ages $\tau$ from Table~2.  The uncertainties in the primary eclipse depths are dominated by systematic errors $\delta \Delta I_1$~$\approx$~0.01~mag, except for the two systems with the deepest eclipses that have $\Delta I_1$ = 1.73\,$\pm$\,0.05~mag~(ID-7842) and $\Delta I_1$~=~2.82\,$\pm$\,0.14~mag~(ID-21452).  The uncertainties in the reflection effect amplitudes are  $\delta \Delta I_{\rm refl}$ $\approx$ 0.003 mag, and the uncertainties in the ages $\tau$ are as those reported in Table 2. 

In Fig.~9,  we compare the eclipsing binary properties listed in Table 4, where we have excluded ID-5898 which is most likely biased toward shallower eclipses and younger ages due to third light contamination.  The empirical properties of primary eclipse depth $\Delta I_1$ and reflection effect amplitude $\Delta I_{\rm refl}$ are positively correlated (Spearman rank correlation coefficient $\rho$~=~0.85) at a statistically significant level (probability of no correlation  $p$ = 2$\times$10$^{-6}$).  This is because both $\Delta I_1$ and  $\Delta I_{\rm refl}$ are inextricably linked to the radius $R_2$ of the pre-MS companion.  The age $\tau$ is anti-correlated with both $\Delta I_1$ and $\Delta I_{\rm refl}$ ($\rho$~=~$-$0.70 and $\rho$~=~$-$0.83, respectively) because older pre-MS stars are systematically smaller.  Although still statistically significant ($p$~=~3$\times$10$^{-6}$\,-\,4$\times$10$^{-4}$), these correlations are not as strong because the radius of a pre-MS star also depends on its mass in addition to its age. 

The three properties $\Delta I_1$, $\Delta I_{\rm refl}$, and $\tau$ of the eclipsing binaries are all significantly correlated with the mean physical radii $\langle r \rangle _{\rm H\,II}$ of the H\,\textsc{ii} regions with which they are associated. Namely, younger eclipsing binaries with deeper primary eclipses and larger reflection effect amplitudes are typically embedded in bright and/or condensed H\,\textsc{ii} regions.  These correlations are statistically significant ($p$ = 5$\times$10$^{-4}$\,-\,0.01), but the mapping between the eclipsing binary properties and the radii of the H\,\textsc{ii} regions are not one-to-one ($|\rho|$~$\approx$~0.6\,-\,0.7).  For example, ID-21452, which happens to be our youngest system ($\tau$~=~0.6\,$\pm$\,0.4\,Myr) with the deepest eclipse ($\Delta I_1$~$\approx$~2.8 mag), resides in the famous, bright, large H\,\textsc{ii} region 30 Doradus (also known as the Tarantula Nebula; RA\,$\approx$\,5$^{\rm h}$39$^{\rm m}$ and DE\,$\approx$\,$-$69.1$^{\rm o}$ in Fig.~8).  Such large, bright H\,\textsc{ii} regions can host multiple episodes of star formation \citep{Crowther2013}.  Specifically, 30~Doradus contains an older population of stars with $\tau$~=~20\,-\,25~Myr, which is consistent with its larger size, and a more recent generation that is $\tau$ $\lesssim$ 1\,-\,2~Myr old, which is consistent with the age derived for ID-21452 \citep{Massey1998,Grebel2000}.

\begin{figure}[t]
\centerline{
\includegraphics[trim = 0.1cm 0.2cm 1.2cm 0.1cm, clip=true, width=3.6in]{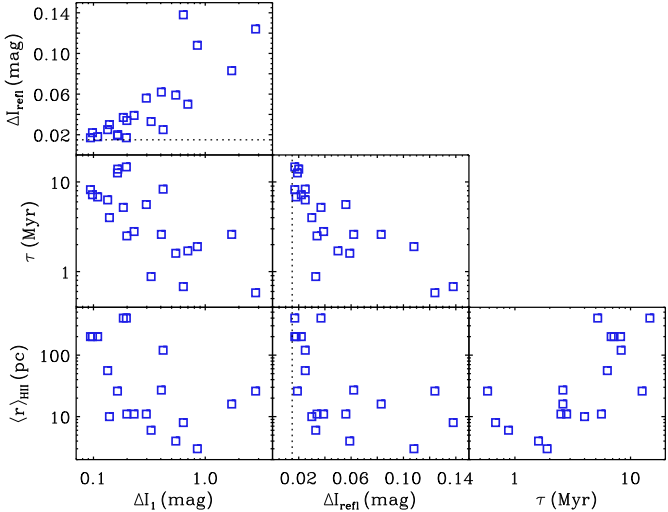}}
\caption{ Properties of eclipsing binaries and H\,\textsc{ii} regions.  The reflection effect amplitudes $\Delta I_{\rm refl}$, primary eclipse depths $\Delta I_1$, ages $\tau$, and physical radii of the H\,\textsc{ii} regions $\langle r \rangle _{\rm H\,II}$ in which the eclipsing binaries reside are all correlated with each other at a statistically significant level. Our eclipsing binaries provide powerful diagnostics and constraints for the dynamical evolution and expansion velocities $\langle v \rangle _{\rm H\,II}$~$\approx$~10\,-\,30~km~s$^{-1}$ of H\,\textsc{ii} regions.  See text for details and a discussion of uncertainties.}
\end{figure}

The properties of our nascent eclipsing binaries provide powerful diagnostics for the long-term evolution of H\,\textsc{ii} regions.  Namely, the mean expansion velocity $\langle v \rangle _{\rm H\,II}$ = $\langle \langle r \rangle _{\rm H\,II} / \tau \rangle$ of H\,\textsc{ii} regions derives from the slope of the observed correlation in the bottom right panel of Fig.~9.  For the 12 bright and centrally condensed  H\,\textsc{ii} regions with $\langle r \rangle _{\rm H\,II}$ = 3\,-\,30~pc, we find a mean expansion velocity of $\langle v \rangle _{\rm H\,II}$ = 8\,$\pm$\,3\,km~s$^{-1}$.  This is consistent with both observed and theoretical estimates of $\langle v \rangle _{\rm H\,II}$~$\approx$~10~km~s$^{-1}$ during the subsonic expansion phase of H\,\textsc{ii} regions when $\tau$ $\approx$ 0.01\,-\,5 Myr \citep{Yorke1986,Cichowolski2009}.  For the seven large and diffuse H\,\textsc{ii} regions with $\langle r \rangle _{\rm H\,II}$~$>$~30~pc, we calculate $\langle v \rangle _{\rm H\,II}$ = 29\,$\pm$\,8\,km~s$^{-1}$.  This coincides with the observed range of expansion velocities $v$ $\approx$ 15\,-\,45 km s$^{-1}$ in giant H\,\textsc{ii} shell-like regions within nearby galaxies \citep{Chu1994,Tomita1998}.  Our ability to measure the ages of several eclipsing binaries to accuracies of $\approx$25\% give tight constraints for the dynamical evolution of the  H\,\textsc{ii} regions in which they formed.

\section{The Intrinsic Close Binary Statistics}

  In the following, we determine the intrinsic fraction~$F$ of B-type MS stars that have close low-mass companions. We utilize the properties and statistics of our nascent eclipsing binaries, and so we must correct for geometrical and evolutionary selection effects in our magnitude-limited sample.  To achieve this, we estimate the probability density functions (\S5.1) for the eight parameters in \S3 that describe our physical models.  With these distributions, we calculate the probability of observing reflection effects using two approaches: a  simple estimate (\S5.2) and a detailed Monte Carlo simulation (\S5.3\,-\,5.4). 

\subsection{Probability Density Functions}

The distribution of dust extinction toward B-type MS stars in the LMC peaks at $A_V$~$\approx$~0.4~mag, i.e. $A_I$~$\approx$~0.2~mag according to our adopted reddening law, with a long tail toward larger values \citep{Zaritsky1999,Zaritsky2004}. To match these observations, we utilize a beta probability distribution to model the extinction distribution in the I-band:

\begin{equation}
  p_{A_I} = 30 A_I (1-A_I)^4~~{\rm for}~0\,<\,A_I\,<\,1, 
\end{equation}

\noindent where $A_I$ is in magnitudes.  The measured dust extinctions $A_I$ of our 22 reflecting eclipsing binaries are also consistent with this distribution (see Table 2 and Fig.~7).  

To quantify the probability density functions for $M_1$ and $\tau$, we estimate the initial mass function (IMF) and recent star-formation history (SFH) within the OGLE-III footprint of the LMC.  We consider a single power-law IMF for massive primaries:

\begin{equation}
 dN = k M_1^{-\alpha} dM_1~~{\rm for}~3\,M_{\odot}\,<\,M_1\,<\,30\,M_{\odot}
\end{equation}

\noindent where the normalization constant $k$ and IMF slope $\alpha$ are free parameters.  Note that $\alpha$ = 2.35 corresponds to the standard Salpeter value. We model the relative SFH of the LMC for ages 0\,Myr~$<$~$\tau$~$<$~320\,Myr, where $\tau$ = 320\,Myr is the MS lifetime of the lowest mass primaries $M_1$ = 3\,\Msun\ we have considered.  We set the relative star-formation rate during recent times 0\,Myr~$<$~$\tau$~$<$~10\,Myr to unity, and consider five free parameters $A$-$E$ to describe the SFH at earlier epochs:

\begin{equation}
  SFH(\tau) =
  \begin{cases}
    ~1~~{\rm for}~~~0\,{\rm Myr}\,\le\,\tau\,<\,~10\,{\rm Myr} \\
    A~~{\rm for}~~10\,{\rm Myr}\,\le\,\tau\,<\,~20\,{\rm Myr}  \\
    B~~{\rm for}~~20\,{\rm Myr}\,\le\,\tau\,<\,~40\,{\rm Myr}  \\
    C~~{\rm for}~~40\,{\rm Myr}\,\le\,\tau\,<\,~80\,{\rm Myr}  \\
    D~~{\rm for}~~80\,{\rm Myr}\,\le\,\tau\,<\,160\,{\rm Myr}  \\
    E~~{\rm for}~160\,{\rm Myr}\,\le\,\tau\,<\,320\,{\rm Myr} \\
   \end{cases} 
\end{equation}

To measure the IMF and SFH model parameters, we utilize the observed present-day luminosity function of MS stars in the OGLE-III LMC database \citep{Udalski2008}.  In Fig.~10, we show the observed magnitude distribution across 15.0 $<$ $\langle I \rangle$ $<$ 18.0 for early-type MS systems with $-$0.25 $<$ $\langle V - I \rangle$ $<$ 0.20.  We have extended our magnitude range to include brighter, short-lived O-type primaries to better constrain the more recent SFH within the OGLE-III LMC footprint. 

 To account for systematic errors caused by unresolved binary stars in the OGLE-III LMC database, we consider two models.  For Model 1, we assume all stars are single, and so the magnitude $\langle I \rangle$ and color $\langle V - I \rangle$ of a system is simply determined by $M_1$, $\tau$, and $A_I$ according to our adopted stellar tracks. For Model~2, we assess the bias in the luminosity distribution due to companions $q$~$\gtrsim$~0.7 that are comparable in mass and luminosity to the primary.  This bias in the luminosity distribution of binary stars was first discussed by \citet{Opik1923}, and we have previously investigated this {\"O}pik effect in the context of stellar populations in extragalactic environments \citep{Moe2013}.  In short, we must approximate the total fraction of B-type MS stars with companions $q$~$\gtrsim$~0.7 across all orbital periods that can measurably affect the luminosity of the system.  For Model 2, we therefore assume a 100\% total binary star fraction and an overall mass-ratio distribution $p_q$~$\propto$~$q^{-0.4}$\,$dq$ across 0.05 $<$ $q$ $<$ 1.0, which is consistent with current observations of B-type MS stars \citep{Kobulnicky2007,Kouwenhoven2007,Rizzuto2013}.  The companion star fraction may exceed 100\% for B-type MS stars, but this is most likely at the expense of increasing the number of low-mass tertiaries that are not easily detectable.  Hence, the fraction of B-type MS primaries that have luminous companions $q$ $\gtrsim$ 0.7 is robust at (23\,$\pm$\,10)\%.   

\begin{figure}[t]
\centerline{
\includegraphics[trim = 0.3cm 0.2cm 1.0cm 0.1cm, clip=true, width=3.3in]{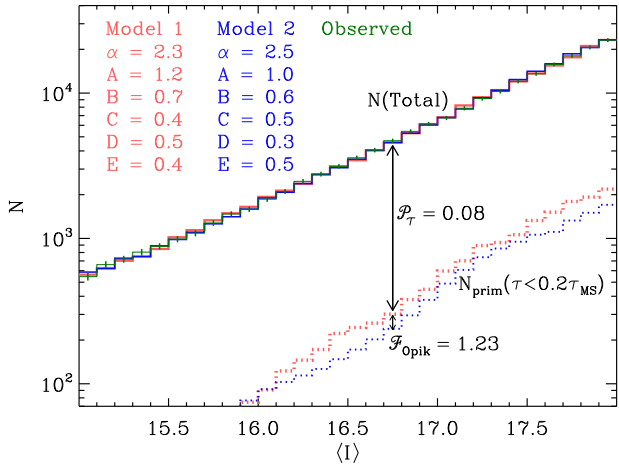}}
\caption{Present-day luminosity function of MS stars in the LMC.  We compare the $\langle I \rangle$ distribution of MS systems with $-$0.25 $<$ $\langle V - I \rangle$ $<$ 0.20 in the OGLE-III LMC database (green) to simulated models assuming all systems are single stars (Model~1~-~red) or binaries (Model~2~-~blue).  We find similar fit parameters between these two models for the slope $\alpha$ of the IMF and relative rates  $A$\,-\,$E$ of star formation.  For our simulated stellar populations, we also display the predicted number of primaries with ages that are $<$20\% their MS lifetimes (dotted).  In a magnitude-limited sample, only ${\cal P}_{\tau}$ $\approx$ 8\% of systems are young enough to have pre-MS companions that are capable of producing detectable reflection effects. In a population of binaries that contain companions $q$~$\gtrsim$~0.7 that are comparable in mass and luminosity to the primary, the total luminosity function is biased toward these bright binaries according to the {\"O}pik effect.  Hence, there are {\it fewer} total primaries by a factor of ${\cal F}_{\rm Opik}$ $=$ 1.23\,$\pm$\,0.10. }
\end{figure}

By implementing a Monte Carlo technique, we generate a population of stars (Model 1) or binaries (Model 2) using our adopted evolutionary stellar tracks and models for the IMF, SFH, and dust extinction distribution.  To constrain the IMF and SFH model parameters,  we minimize the $\chi^2$ statistic between the observed and simulated present-day $\langle I \rangle$ distributions (see Fig.~10).  For both models, we measure a primary star IMF that is consistent with the Salpeter value.  We also find that the star-formation rate has been relatively constant over the past $\approx$\,20 Myr, but was $\approx$40\% the present-day value at earlier epochs $\tau$ $>$ 80 Myr.  This is consistent with other measurements of the SFH in the LMC \citep{Harris2009,Indu2011}.   The uncertainties in the overall binary properties have little influence on our derived slope of the IMF or the relative SFH.  We therefore adopt parameters between our two models, namely $\alpha$ = 2.4, $A$ = 1.1, $B$~=~0.7, $C$ = 0.5, and $D$=$E$=0.4. 

 Only the normalization constant between our single and binary star populations significantly differ.  For our binary population, we measure $\approx$20\% fewer systems due to the {\"O}pik effect.  We also find 20\% more total mass in our binary population because the average binary contains $\approx$1.4 times the mass of the primary, i.e. $\langle q \rangle$ $\approx$ 0.4.  When we generate synthetic eclipsing binary light curves (\S5.3), we simulate only systems that are similar to our 22 observed eclipsing binaries.  Quantitatively,  we generate only B-type MS stars with low-mass companions $q$ = 0.06\,-\,0.40 at short orbital periods $P$~=~3.0\,-\,8.5~days. We therefore need to correct for the luminosity bias of $q$~$\gtrsim$~0.7 companions. Because the observed luminosity distribution is biased toward these binaries with equally bright components, the distribution is biased {\it against} single stars as well as binaries with faint, low-mass companions $q$ $\lesssim$ 0.7.  We therefore multiply our calculated intrinsic fraction $F$ of low-mass companions by a correction factor of ${\cal F}_{\rm Opik}$~=~1.23\,$\pm$\,0.10 to account for this  {\"O}pik effect.

The probability density functions for the remaining physical model parameters are easier to quantify.  We assume random epochs of primary eclipse minima $t_{\rm o}$.  We also assume random orbital orientations so that cos($i$)~=~[0,\,1] is uniformly distributed on this interval.  We select secondary albedos from a uniform distribution across the interval $A_2$~=~[0.3,\,1.0], which encompasses the range of observed albedos in our eclipsing binaries with reflection effects (see Table~3). Although the average albedo of this distribution $\langle A_2 \rangle$~=~0.65 is slightly lower than the observed average $\langle A_2 \rangle$~$\approx$~0.71, the latter is a posterior average and companions with higher albedos are more likely to be detected.  Also, the albedo~$A_2$ may be dependent on the effective temperature~$T_2$ \citep{Claret2001}, but small correlations between model parameters are second-order effects in our overall calculations. We assume log~$P$ is uniformly distributed across the interval $P$~=~3.0\,-\,8.5~days, which is consistent with observations of binaries with B-type MS primaries \citep{Abt1990,Kobulnicky2007,Kouwenhoven2007}.  Reasonable deviations from this distribution have little effect on our statistics, especially considering we are examining such a narrow window of orbital periods.  Finally, in order to calculate the detectability of reflection effects as a function of mass ratio, we consider four logarithmic $d$log$q$~=~0.2 intervals across the total range log~$q$~=~$-$1.2\,-\,$-$0.4, i.e. $q$~=~0.06\,-\,0.40. In our detailed Monte Carlo simulations, we treat each of these four mass-ratio bins independently, and select log~$q$ from a uniform distribution within each interval.  Again, the precise distribution of mass ratios within each narrowly divided bin is inconsequential to our overall uncertainties.  

\subsection{Simple Estimate}

Before we utilize a Monte Carlo technique to generate synthetic light curves for a population of eclipsing binaries, we first perform a simple calculation.  Using the measured properties of our 22 eclipsing binaries, we estimate the probability ${\cal P}_{\rm refl}$ that a B-type MS primary and low-mass companion have the necessary configuration to produce observable eclipses and reflection effects. 
In this simple estimate, we do not account for all eight physical model parameters outlined above.  Instead, we consider only the following three main selection effects.  

 First, eclipsing binaries must have nearly edge-on orientations so that the eclipses are deep enough to be observed given the sensitivity of the OGLE-III LMC observations.  The observed eclipsing binaries with reflection effects in our sample generally have $i$~$\gtrsim$~80$^{\rm o}$ (Table~2).  This implies the probability of having sufficiently edge-on inclinations is ${\cal P}_i$ $\approx$ cos(80$^{\rm o}$) $\approx$ 0.17.  

Second, the observed eclipsing binaries generally have short orbital periods.  This is not only due to geometrical selection effects, but also because irradiation effects quickly diminish with orbital separation.  The majority of our systems have $P$~$=$~3.0\,$\mbox{-}$\,5.5~days, implying ${\cal P}_P$~$\approx$~(log~5.5~$-$~log~3.0)/(log~8.5~$-$~log~3.0)~$\approx$~0.6 if the intrinsic distribution of orbital periods is uniform in log\,$P$. 

Finally, our reflecting eclipsing binaries must be young enough so that the companion is still on the pre-MS, but bright enough to be contained in our magnitude-limited sample.  A B-type MS primary can have a pre-MS companion only if the age of the binary $\tau$ is a certain fraction of the primary's MS lifetime $\tau_{\rm MS}$.  For moderate mass ratios $q$~$\approx$~0.25, the ages must be $\tau$~$\lesssim$~0.1~$\tau_{\rm MS}$.  For binaries with extreme mass ratios $q$~$\approx$~0.15, close orbits $P$~=~3\,-\,4~days, and bright massive primaries $M_1$~=~12\,-\,16~\Msun, we can discern reflection effects up to $\tau$~$\approx$~0.5\,$\tau_{\rm MS}$ (Table 3, Fig.~4, and left panel of Fig.~11).  For our simple estimation purposes, we adopt an average criterion that the binary must have an age $\tau$~$<$~0.2\,$\tau_{\rm MS}$ in order for the companion to be a pre-MS star.  One may initially assume that the probability of observing such a young binary is ${\cal P}_{\tau}$ = 0.2, but this is not the case for our magnitude-limited sample. By incorporating our simulated stellar populations used to quantify the SFH and IMF above, we display in Fig.~10 the number of systems with primaries that have ages $\tau$ $<$ 0.2 $\tau_{\rm MS}$ for each magnitude bin.  The true probability that a system has such a young age is ${\cal P}_{\tau}$~$=$~${\cal N}$($\tau$\,$<$\,$0.2\tau_{\rm MS}$)/${\cal N}$(total)~$\approx$~0.08, which is a factor of 2\,-\,3 times lower than the crude estimate of ${\cal P}_{\tau}$ = 0.2.  Late-B MS primaries with $M_1$~$\approx$~3\,-\,6~\Msun\ can have observed magnitudes $\langle I \rangle$ $<$ 18.0 only if they are older and more luminous on the upper MS.  Alternatively, in order to see a system with $\tau$ $<$ 0.2$\tau_{\rm MS}$ and $\langle I \rangle$~$<$~18.0, the primary must be rather massive with $M_1$ $\gtrsim$ 6\,\Msun.  Note that all of our observed eclipsing binaries with reflection effects have early B-type MS primaries with $M_1$~$\gtrsim$~6\,\Msun.  Because the IMF is significantly weighted toward lower-mass primaries, our magnitude-limited sample is dominated by late-B primaries that are systematically older on the upper MS.  This is the reason why the probability of observing a young system with $\tau$ $<$ 0.2\,$\tau_{\rm MS}$ in our magnitude-limited sample is only ${\cal P}_{\tau}$ $\approx$ 0.08.

Putting these three factors together, then the probability of observing reflection effects is ${\cal P}_{\rm refl}$~=~${\cal P}_i$\,${\cal P}_{\tau}$\,${\cal P}_P$~$\approx$~0.8\%.  In our actual sample, we selected ${\cal N}_{\rm B}$ = 174,000 B-type MS stars from the OGLE-III LMC survey.  From this population, we observed  ${\cal N}_{\rm obs}$ = 19 eclipsing binaries that exhibit reflection effects with $q$ = 0.06\,-\,0.25 companions and $P$ = 3.0\,-\,8.5 days.   After accounting for the correction factor ${\cal F}_{\rm Opik}$ = 1.23 due to the  {\"O}pik effect, then the intrinsic fraction of B-type MS stars with low-mass $q$ = 0.06\,-\,0.25 companions and short orbital periods $P$~=~3.0\,$\mbox{-}$\,8.5~days is $F$\,$=$\,(${\cal N}_{\rm obs}\,{\cal F}_{\rm Opik}$)\,/\,(${\cal N}_{\rm B}\,{\cal P}_{\rm refl}$)\,$\approx$ (19\,$\times$\,1.23)\,/\,(174,000\,$\times$\,0.008)\,$\approx$\,1.7\%. This is only an approximation as we need to quantify ${\cal P}_{\rm refl}$ as a function of $q$ in a more robust manner.  Nonetheless, this simple analysis separates the individual selection effects and illustrates the difficulty in detecting young, low-mass pre-MS companions that eclipse B-type MS stars.  

\begin{figure*}[t]
\centerline{
\includegraphics[trim = 0cm 3.1cm 0cm 2.8cm, clip=true, width=6.55in]{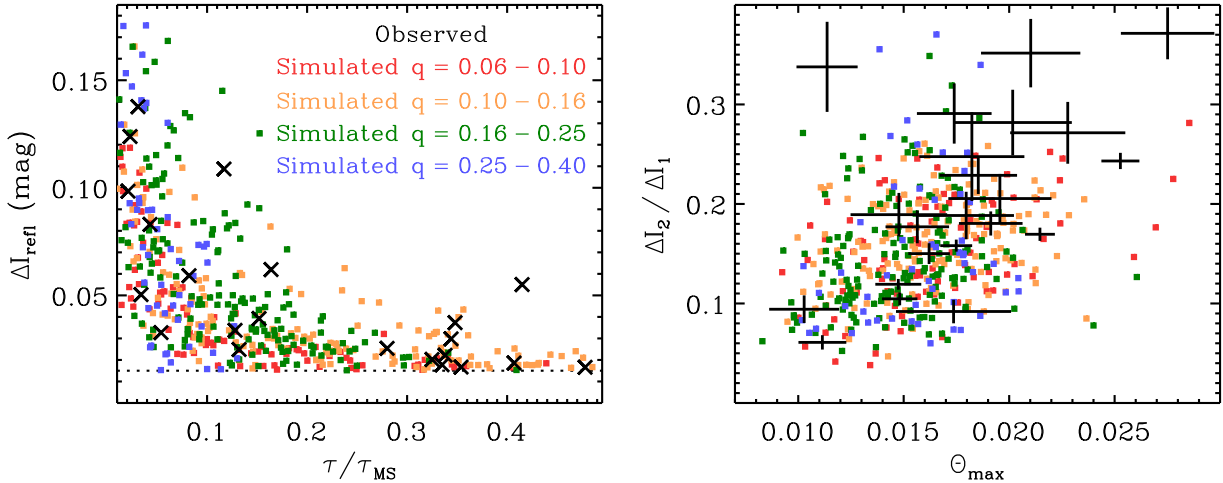}}
\caption{We compare the properties of the 22 observed eclipsing binaries (black) to the 468 simulated systems (color) with B-type MS primaries, low-mass $q$ = 0.06\,-\,0.40 pre-MS companions, and pronounced reflection effects as listed in Table 5.  Left panel: The anti-correlation between the reflection effect amplitude and the age relative to the MS lifetime of the primary is similar to the trend seen in the middle panel of Fig 9.  Binaries with moderate mass ratios $q$ = 0.25\,-\,0.40 (blue) can have pre-MS companions only at extremely young ages $\tau$ $\lesssim$ 0.1 $\tau_{\rm MS}$, while lower-mass companions can have pre-MS evolutions that last up to $\approx$50\% the MS lifetime of the primary.  Right panel: Identical parameter space used in Fig.~3 to identify our systems and differentiate them from other classes of eclipsing binaries.  Our simulated systems correspond well to the observed population.  We can therefore easily identify eclipsing binaries with B-type MS primaries and low-mass pre-MS companions at $P$ = 3.0\,-\,8.5 days by selecting systems with reflection effect amplitudes $\Delta I_{\rm refl}$~$>$~0.015~mag, maximum eclipse widths $\Theta_{\rm max}$ $<$ 0.03, and ratios of eclipse depths $\Delta I_2$/$\Delta I_1$ $<$ 0.4. }
\end{figure*}

\subsection{Detailed Monte Carlo Simulation}

We now perform a more detailed Monte Carlo simulation by synthesizing \textsc{Nightfall} light curves for a population of eclipsing binaries. Using our probability density functions, we select a binary with a primary mass $M_1$, mass ratio $q$, age $\tau$, and dust extinction $A_I$.  Based on our adopted evolutionary stellar tracks, we then determine the observed magnitude $\langle I \rangle$ and color $\langle V - I \rangle$ of the binary.  If the magnitudes and colors do not satisfy our photometric selection criteria, we generate a new binary.  Otherwise, we count its contribution toward our statistics of binaries with B-type MS primaries.  For each of the four mass-ratio bins, we simulate ${\cal N}_{\rm sim}$~=~2$\times$10$^{4}$ binaries that satisfy our magnitude and color criteria and therefore have B-type MS primaries.

 For each of these binaries, we then select $P$, $t_o$, $i$, and $A_2$ according to their respective probability density functions.  To be detectable as a detached, closely orbiting eclipsing binary, the system must be old enough so that the pre-MS companion neither fills its Roche~lobe nor accretes from a thick circumstellar disk (see \S2).  Based on our observed systems (Table 2), we therefore require the companion Roche-lobe fill-factor to be $RLFF_2$~$<$~80\% and the age $\tau$~$>$~0.5 Myr.  If these criteria are satisfied, we synthesize an eclipsing binary light curve with \textsc{Nightfall} as in \S3.2 according to our eight randomly generated physical model parameters.  We note that we have implicitly assumed that the orbital periods of binaries do not significantly change between $\tau$~=~0.5~Myr and the time $\tau$ $\approx$ $\tau_{\rm MS}$ $\approx$ 25 Myr until the primary fills its Roche lobe.  It is possible, however, that subsequent dynamical interactions with a tertiary can harden the orbit of the inner binary \citep{Kiseleva1998,Naoz2014}. This may  bring additional systems into our parameter space $P$~=~3.0\,-\,8.5~days after the secondary has contracted into a MS star.  Hence, we can only measure the fraction of B-type MS stars with close, low-mass companions at 0.5~Myr~$\lesssim$~$\tau$~$\lesssim$~0.4~$\tau_{\rm MS}$~$\approx$~10 Myr.  The binary fraction at earlier or later epochs may be different.

We now ensure our synthesized light curve matches the cadence and precision of the OGLE-III LMC observations.  We therefore interpolate our theoretical \textsc{Nightfall} eclipsing binary light curve at ${\cal N}_I$ = 470 randomly selected orbital phases. Obviously, the total photometric errors increase toward fainter systems.  We relate the I-band photometric error to the I-band magnitude according to the following:

\begin{equation}
 \sigma_{\rm fit}(I)  = \big[1+10^{(I-17.0)/2}\big] \times 0.0072\,{\rm mag}.
\end{equation}

\noindent This simple formula fits the observed rms scatter in the eclipsing binary light curves as discussed in \S2.1 (see black curve in Fig.~2).  For each I-band value in our synthesized light curve, we add random Gaussian noise according to Eq. 10.  

We then fit our analytic model of Gaussians and sinusoids  (Eq. 3) to this simulated I-band light curve by implementing the same Levenberg-Marquardt technique in \S2.1.  To ensure automated and fast convergence toward the true solution, we choose initial model parameters motivated by the properties of the eclipsing binary.  For example, because we only synthesize eclipsing binaries with circular orbits in our Monte Carlo simulations, we select $\Phi_2$ = 0.5 as the initial estimates in our analytic models.  In this manner, for each synthetic eclipsing binary generated by \textsc{Nightfall}, we measure the analytic model parameters, e.g. $\Delta I_1$, $\Theta_1$, $\Delta I_{\rm refl}$, etc., and their respective errors. 

 To be considered an eclipsing binary with observable reflection effects, we impose the same selection criteria as in \S2. The reflection effect amplitude must be $\Delta I_{\rm refl}$~$>$~0.015~mag with a 1$\sigma$ error that is $<$20\% the actual value.  We require the uncertainties in the eclipse depths $\Delta I_1$ and $\Delta I_2$ and eclipse widths $\Theta_1$ and $\Theta_2$ to be $<$25\% their respective values.  The full light curve amplitude $\Delta I$~=~$\Delta I_1$~+~$\Delta I_{\rm refl}$ must be deep enough to be detectable by the OGLE-III LMC survey according to Eq. 6. Finally, the maximum eclipse width $\Theta_{\rm max}$~$<$~0.03 and ratio of eclipse depths $\Delta I_2$/$\Delta I_1$~$<$~0.4 need to satisfy our selected parameter space as shown in Fig.~3.  If the synthetic light curve satisfies all these properties, then it contributes toward the number  ${\cal N}_{\rm refl}$ of eclipsing binaries with reflection effects. 

\subsection{Results}

We generated a total of 4\,$\times$\,${\cal N}_{\rm sim}$ = 8$\times10^4$ eclipsing binaries with B-type MS primaries, low-mass companions $q$~=~0.06\,-\,0.40, and magnitudes and colors that satisfy our photometric selection criteria.  Of these simulated systems, only ${\cal N}_{\rm refl}$ = 468 eclipsing binaries have the necessary ages and orientations to produce detectable reflection effects and eclipses.  We compare the properties of these ${\cal N}_{\rm refl}$ = 468 eclipsing binaries from our Monte Carlo simulations to our 22 observed systems in Fig.~11.  In the left panel, we can see that larger reflection effect amplitudes dictate younger relative ages $\tau$/$\tau_{\rm MS}$ for both the observed and simulated populations.  The only system that noticeably deviates from this trend is ID-18330, which has a moderate reflection effect $\Delta I_{\rm refl}$~=~0.056 and older relative age $\tau$/$\tau_{\rm MS}$~=~0.42.  We note that ID-18330 has a large intrinsic V-band scatter $f_{\sigma, V}$~=~3.0 and a modest fit statistic $\chi^2$/$\nu$~=~1.16, so that the systematic error in our measured age for this system is larger than usual. In any case, ID-18330 is relatively bright $\langle I \rangle$~=~16.1, which requires a massive primary $M_1$~$\approx$~15\,\Msun.  Hence, ID-18330 has a large relative age $\tau$/$\tau_{\rm MS}$~=~0.42 mainly because the massive primary is short-lived with $\tau_{\rm MS}$~$\approx$~13\,Myr. We therefore expect the majority of eclipsing binaries with $\tau$/$\tau_{\rm MS}$~$>$~0.2 to have $\Delta I_{\rm refl}$ $<$ 0.04~mag, while only the few systems with short-lived, massive primaries $M_1$~$\gtrsim$~14\,\Msun\ can have $\Delta I_{\rm refl}$~$\approx$~0.04\,-\,0.06 mag at these older relative ages.

In the right panel of Fig.~11, we compare the maximum eclipse depths $\Theta_{\rm max}$ versus the ratio of eclipse depths $\Delta I_2/\Delta I_1$ for our 468 simulated and 22 observed eclipsing binaries.  Both the observed and simulated systems cluster near $\Theta_{\rm max}$ = 0.017 and $\Delta I_2$/$\Delta I_1$~=~0.2. As discussed in \S2, the pre-MS companions are detached from their Roche lobes and have low luminosities, which require $\Theta_{\rm max}$~$<$~0.3 and $\Delta I_2$/$\Delta I_1$ $<$ 0.4, respectively.  Only the three systems at the top with  0.3~$<$~$\Delta I_2$/$\Delta I_1$~$<$~0.4 are marginally discrepant with the simulated population.  Two of these, ID-1965 and ID-6630, have reflection effect amplitudes $\Delta I_{\rm refl}$ = 0.017\,-\,0.019 mag just above our detection limit of $\Delta I_{\rm refl}$ = 0.015 mag  and companion properties that are consistent with the zero-age MS (see Fig.~6).  If we had assumed a detection limit of $\Delta I_{\rm refl}$ = 0.012 mag in our Monte Carlo simulations, we would have synthesized an additional $\approx$15\% of reflecting eclipsing binaries. The majority of these additional systems would have occupied the upper-right portion in the right panel of Fig.~11, consistent with ID-1965 and ID-6630.  The other eclipsing binary, ID-17217,  that is slightly discrepant with the simulated population has a slightly asymmetric light curve profile between eclipses (see Fig.~5).  This asymmetry is most likely due to an eccentric orbit, as indicated by the phase of the secondary eclipse $\Phi_2$ = 0.490.  However, the slight asymmetry could also be caused by a disk or hot spot, similar to other systems we observed with $\Delta I_2$/$\Delta I_1$~$>$~0.4 (see Fig.~3).  Even if this one system is a contaminant in our sample, it has a negligible effect on our statistics.  Most importantly, the 19 observed eclipsing binaries with $\Delta I_2$/$\Delta I_1$ $<$ 0.3 match the simulated population and clearly have young, low-mass companions.

\begin{table}[t]\footnotesize
\begin{flushleft}
{\small {\bf Table 5.} Results of Monte Carlo simulation.}
\end{flushleft}
\vspace{-0.3cm}
\renewcommand{\tabcolsep}{4.0pt}
\centerline{
\begin{tabular}{|c|c|r|r|c|c|}
\hline
   log\,$q$             &      $q$            & ${\cal N}_{\rm obs}$ & ${\cal N}_{\rm refl}$  &  ${\cal P}_{\rm refl}$\,(\%)  &  $F$\,(\%)  \\
\hline
 $-$1.2\,-\,$-$1.0      & 0.06\,-\,0.10       &      5~             &      92~              &     0.46         &         0.77\,$\pm$\,0.34    \\
\hline
 $-$1.0\,-\,$-$0.8      & 0.10\,-\,0.16       &      8~             &     156~              &     0.78         &         0.73\,$\pm$\,0.26    \\
\hline
 $-$0.8\,-\,$-$0.6      & 0.16\,-\,0.25       &      6~             &     167~              &     0.84         &         0.51\,$\pm$\,0.21    \\
\hline
 $-$0.6\,-\,$-$0.4      & 0.25\,-\,0.40       &      2~             &      53~              &     0.26         &         0.53\,$\pm$\,0.37    \\
\hline
{\bf $-$1.2\,-\,$-$0.6} & {\bf 0.06\,-\,0.25} &   \,{\bf 19}~       &    {\bf 415}~         &     {\bf 0.69}   &    {\bf 2.0\,$\pm$\,0.6}    \\
\hline
\end{tabular}}
\end{table}

We present the statistics of our Monte Carlo simulations in Table 5.  For each of our four mass-ratio intervals, we report the number ${\cal N}_{\rm obs}$ of observed systems in our sample, the number ${\cal N}_{\rm refl}$ of simulated systems that exhibit reflection effects, the probability of observing reflection effects ${\cal P}_{\rm refl}$ = ${\cal N}_{\rm refl}$/${\cal N}_{\rm sim}$ where ${\cal N}_{\rm sim}$ = 2$\times$10$^4$, and the intrinsic binary fraction $F$\,$=$\,(${\cal N}_{\rm obs}\,{\cal F}_{\rm Opik}$)\,/\,(${\cal N}_{\rm B}\,{\cal P}_{\rm refl}$) where ${\cal F}_{\rm Opik}$ = 1.23 and ${\cal N}_{\rm B}$ = 174,000.  As expected, ${\cal P}_{\rm refl}$~$\approx$~0.8\% is largest for systems with $q$ = 0.10\,-\,0.25.  The probability ${\cal P}_{\rm refl}$ $\approx$ 0.3\% quickly diminishes toward larger mass ratios because the pre-MS timescales of more massive companions are markedly shorter.  Even though lower mass companions $q$ = 0.06\,-\,0.10 have longer pre-MS evolutions, the probability ${\cal P}_{\rm refl}$ $\approx$ 0.5\% of observing eclipses and reflection effects is low because the radii of the companions are systematically smaller (see Figs.~4 and~6).  

In the bottom row of Table 5, we combine the statistics for our three smallest mass-ratio bins.  For our observed sample of 19 eclipsing binaries in this interval, the relative error from Poisson statistics is 23\%.  We expect a systematic error of 15\% due to uncertainties in our light curve modeling.  For example, the few systems with $q$ = 0.20\,-\,0.25 could easily shift toward solutions with $q$~$>$~0.25 outside our defined interval of extreme mass ratios.  We also estimate a 10\% systematic error due to third light contamination and the possibility of mimics in our sample.  For example, ID-5898 may have $q$ $<$ 0.25 and should therefore be added to our statistics (see \S3), while ID-17217 may host a disk and/or hot spot and therefore should be removed from our statistics (see above).  Finally, the correction factor ${\cal F}_{\rm Opik}$ due to the {\"O}pik effect is uncertain by 10\%.  We add all these sources of error in quadrature, and find the total relative error in our binary statistics is $\approx$30\%.  Therefore, $F$~=~(2.0\,$\pm$\,0.6)\% of young B-type MS stars have low-mass companions $q$ = 0.06\,-\,0.25 with short orbital periods $P$ = 3.0\,-\,8.5 days.  This result from our detailed Monte Carlo simulation is consistent with our simple estimate of $F$ $\approx$ 1.7\%.  The selection effects are therefore well understood and the probability of observing reflection effects is robust.

\section{Discussion}

\subsection{Binary Statistics}
The close binary fraction of MS stars has long been understood to increase with primary mass \citep{Abt1983,Duquennoy1991,Raghavan2010,Sana2012,Duchene2013}.  This correlation between the close binary fraction and spectral type has been primarily based on observations of moderate mass-ratio companions with $q$ $\gtrsim$ 0.25.  In Fig.~12, we show the binary star fraction across orbital periods $P$~=~3.0\,-\,8.5~days as a function of mass ratio $q$ for solar-type primaries \citep{Grether2006}, B-type primaries \citep{Wolff1978,Levato1987,Abt1990}, and O-type primaries \citep{Sana2012}. About 1.0\% of solar-type stars have companions with moderate mass ratios $q$~$>$~0.25 and short orbital periods $P$ = 3.0\,-\,8.5 days.  This increases to $\approx$\,3.8\% for B-type MS stars, and up to $\approx$\,14\% for O-type stars.  Hence, the close binary fraction at moderate mass ratios $q$~$>$~0.25 increases by a factor of $\approx$\,4 between $M_1$~$\approx$~1\,\Msun\ solar-type primaries and $M_1$~$\approx$~10\,\Msun\ B-type MS primaries.  

  As discussed in \S1, SB1s with early-type MS primaries may have companions that are evolved stellar remnants \citep{Wolff1978,Garmany1980}.  We can therefore not reliably infer the frequency of extreme mass-ratio stellar companions from early-type spectroscopic binaries.  The companions in our reflecting eclipsing binaries are unambiguously low-mass, unevolved, pre-MS stars.  After correcting for geometrical and evolutionary selection effects (\S5), we found that (2.0\,$\pm$\,0.6)\% of young B-type MS stars have companions with $q$ = 0.06\,-\,0.25 and $P$ = 3.0\,-\,8.5 days.  Considering 3.8\% of B-type MS stars have companions with $q$~$>$~0.25 across the same period range, then extreme mass-ratio companions $q$~=~0.06\,-\,0.25 constitute one-third of close stellar companions to B-type MS stars. This result indicates the majority of SB1s with B-type MS primaries contain low-mass stellar companions.  This is in disagreement with \citet{Wolff1978}, who suggested SB1s with late-B MS primaries most likely contain white dwarf companions.

\begin{figure}[t]
\centerline{
\includegraphics[width=3.5in]{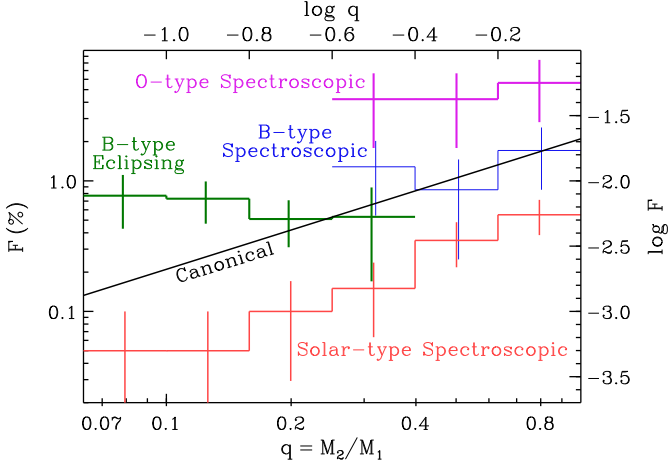}}
\caption{The fraction $F$ of MS primaries that have stellar companions with orbital periods $P$ = 3.0\,-\,8.5 days divided into $d$log$q$ = 0.2 intervals.  In a spectroscopic survey of 71 O-type stars \citep[magenta;][]{Sana2012} and combined sample of 234 B-type stars \citep[blue;][]{Wolff1978,Levato1987,Abt1990}, 10 (14\%) and 9 (3.8\%), respectively, were identified as double-lined spectroscopic binaries in our period range with dynamically measured mass ratios $q$ $>$ 0.25.  Utilizing  eclipsing binaries is the only way of accurately measuring the intrinsic frequency of low-mass unevolved stellar companions to B-type MS stars (green).  In a survey of 2,001 solar-type primaries \citep{Grether2006}, only 25 (1.2\%) were found to be spectroscopic binaries in our period range with mass functions that indicate a stellar secondary companion with $q$ $>$ 0.08 (red).  Population synthesis studies of close binaries canonically assume a uniform mass-ratio distribution \citep{Kiel2006,Ruiter2011,Claeys2014}, i.e. flat in {\it linear} $q$, according to $dF$ $=$ 0.1\,$dq$\,$d$log$P$ (black).}
\end{figure}

 For solar-type MS primaries $M_1$ $\approx$ 1\,\Msun, low-mass companions $M_2$ $\approx$ 0.1\,-\,0.2\,\Msun\ are almost certainly low-mass M-dwarfs \citep{Duquennoy1991,Halbwachs2000}.  In a sample of 2,001 spectroscopic binaries with solar-type MS primaries \citep{Grether2006}, only 4 (0.2\%) had companions with $P$~=~3.0\,-\,8.5 days and $q$~$\approx$~0.08\,-\,0.25.  We found that (2.0\,$\pm$\,0.6)\% of B-type MS stars have stellar companions across the same mass-ratio and period interval, which is a factor of $\approx$10 times larger (Fig.~12).  The frequency of close, extreme-mass ratio companions increases with primary mass even more dramatically than the overall close binary fraction.  

 We can also interpret this trend according to differences in the intrinsic mass-ratio probability distribution {\bf $p_q$}. The mass-ratio distribution is typically described by a power-law  {\bf $p_q$} $\propto$ $q^{\gamma}$\,$dq$.  For close companions to solar-type MS stars, the mass-ratio distribution across 0.08 $<$ $q$ $<$ 1.0 is close to uniform, i.e. $\gamma$ $=$ 0.1\,$\pm$\,0.2  \citep{Grether2006,Raghavan2010}.  By combining the statistics of eclipsing binaries and SB2s with B-type MS primaries,  we measure  $\gamma$~=~$-$0.7\,$\pm$\,0.3 across the broad interval 0.07~$<$~$q$~$<$~1.0 (Fig.~12).  This is consistent with our previous measurement of $\gamma$~=~$-$0.8\,$\pm$\,0.3 in \citet{Moe2013} for close companions ($P$~=~2\,-\,20~days) to B-type MS stars.  In \citet{Moe2013}, however, we used only the primary eclipse depth distribution of eclipsing binaries to recover the intrinsic mass-ratio distribution.  In the present study, we have directly measured the physical properties of companions with extreme mass ratios $q$ = 0.07\,-\,0.25.   Not only does the close binary fraction increase with primary mass, but the mass-ratio distribution also becomes weighted toward smaller values \citep[see also][and references therein]{Duchene2013}.

\subsection{Binary Formation}

 The dearth of short-period, low-mass companions to solar-type MS stars has been investigated in previous spectroscopic binary surveys \citep{Duquennoy1991,Halbwachs2003,Raghavan2010}.  In fact, there appears to be a complete absence of close $q$~$\approx$~0.02\,-\,0.08 companions to solar-type MS stars, commonly known as the brown dwarf desert \citep{Halbwachs2000,Grether2006}.   This is most likely because such low-mass companions would have migrated inward during their formation in the circumstellar disk and subsequently merged with the primary \citep{Armitage2002}.  For luminous and massive B-type MS primaries, however, the circumstellar disk quickly photoevaporates within $\tau$~$\lesssim$~0.3~Myr \citep{AlonsoAlbi2009}. Moreover, B-type MS stars with $q$~$\approx$~0.1 companions have $\approx$10 times more mass and orbital angular momenta than their solar-type counterparts.  Our nascent eclipsing binaries demonstrate that the rapid disk photoevaporation timescales and larger orbital angular momenta of more massive binaries can allow an extreme mass-ratio system to stabilize into a short orbit without necessarily merging.  

As discussed in \S1, there is a body of work indicating that components in close binaries coevolved via fragmentation and competitive accretion in the circumbinary disk \citep{Bate1997,Bate2002,Bonnell2005,Kratter2006}.  Coevolution preferentially leads to binary component masses that are correlated.  The rapid disk photoevaporation timescales around more massive stars suggest competitive accretion may be less significant. It is therefore plausible that less efficient competitive accretion in early-type systems can naturally produce close binaries with extreme mass ratios. This would be consistent with the measured mass-ratio distribution of close early-type binaries, which favors extreme mass ratios more readily than that observed for solar-type binaries.  

 It is also possible that extreme mass-ratio binaries require a different formation mechanism.  The low-mass pre-MS companions in our eclipsing binaries are quite large with moderate Roche-lobe fill-factors 30\%~$<$~$RLFF_2$~$<$~80\%.  Tidal dissipation of orbital energy and angular momentum in a pre-MS star with a large convective envelope is orders of magnitude more efficient than in a MS star \citep{Zahn1989}. Binary formation via tidal capture of low-mass companions  may be substantially more efficient than previously realized \citep{Press1977,Moeckel2007} if one accounts for the long pre-MS timescales of the low-mass secondaries.    Additionally, the pre-MS companions may have been captured with the assistance of dynamical perturbations from an outer tertiary via Kozai cycles and tidal friction \citep{Kiseleva1998,Naoz2014}.  In any case, future formation models of massive stars and close binaries must readily produce these kinds of systems on rapid timescales.

\subsection{Binary Evolution}

Given the short orbital periods $P$ $<$ 10 days of our 22 systems, we expect these binaries will eventually coalesce as the primary evolves off the MS. Low-mass X-ray binaries and millisecond pulsars that form in the galactic field \citep{Kalogera1998,Kiel2006} as well as Type Ia supernovae that explode in elliptical galaxies \citep{Whelan1973,Ruiter2011} can derive from B-type MS primaries with low-mass companions $q$ $\approx$ 0.1\,-\,0.3 at slightly longer orbital periods $P$~$\approx$~10$^2$\,-\,10$^3$~days. These binary population synthesis studies canonically assume a uniform mass-ratio distribution, i.e. $\gamma$~=~0, normalized to 0.1 companions per primary per decade of orbital period (Fig.~12).  We have shown that low-mass companions $q$ $<$ 0.25 to B-type MS stars at short orbital periods $P$~$<$~10 days not only survive, but are found in abundance and constitute one-third of such close companions (i.e., $\gamma$~$\approx$~$-$0.7).  

Photometrically resolved companions to early-type MS stars with $P$~$\gtrsim$~10$^5$~days are generally weighted toward even smaller mass ratios \citep{Preibisch1999,Shatsky2002,Peter2012}.  These wide companions to early-type stars may have formed relatively independently from the primaries, and may therefore have a mass-ratio distribution exponent $\gamma$~=~$-$2.35 that is consistent with random pairings from a Salpeter IMF \citep{Abt1990,Duchene2001}.

 If we interpolate between these two regimes, then we may expect low-mass companions to early-type stars to be plentiful at moderate orbital periods.  Hence, there may be more progenitors of low-mass X-ray binaries, millisecond pulsars, and Type~Ia supernovae than originally assumed. We intend to confirm this conjecture by investigating the properties of massive binaries at intermediate orbital separations.   Specifically, we are in the process of characterizing OGLE-III LMC eclipsing binaries with B-type MS primaries and $P$~$>$~20~days (Moe et al., in prep.). 

\section{Summary}

\begin{enumerate}[leftmargin=*]

\item  {\it New Class of Eclipsing Binaries.}  We analyzed the light curves of 2,206 systems in the OGLE-III LMC eclipsing binary catalog \citep{Graczyk2011} with B-type MS primaries and orbital periods $P$~=~3\,-\,15~days~(\S2.1).  We discovered a subset of 22 detached eclipsing binaries  with short orbital periods ($P$~=~3.0\,-\,8.5~days) that exhibit substantial reflection effects ($\Delta I_{\rm refl}$~$=$~0.017\,-\,0.138~mag) and moderate to deep primary eclipses ($\Delta I_1$~=~0.09\,-\,2.8 mag).  Because such deep eclipses and prominent reflection effects require the secondaries to be comparable in size to the primaries ($R_2$/$R_1$ $>$ 0.3) but markedly cooler ($T_2$/$T_1$~$<$~0.4), we concluded the companions in these 22 eclipsing binaries are large, cool, low-mass pre-MS stars (\S2.2).  

~~Similar irradiation effects have been observed in evolved binaries that contain a hot, low-luminosity, compact remnant in an extremely short orbit ($P$~$\lesssim$~1~day) with a late-type MS companion (\S2.3.1).    Previous observations of young MS\,+\,pre-MS eclipsing binaries have been limited to large mass ratios $q$~$\gtrsim$~0.5, low-mass primaries $M_1$~$\lesssim$~3\,\Msun, and/or systems that are still accreting from a circumbinary disk (\S2.3.2).  Hence, our 22 eclipsing binaries constitute a new class of nascent eclipsing binaries in which a detached, non-accreting, low-mass pre-MS companion discernibly reflects much of the light it intercepts from the B-type MS primary.  We have not yet observed the precise counterparts to these systems in our own Milky Way galaxy, primarily because our sample of continuously monitored ${\cal N}_{\rm B}$ = 174,000 B-type MS stars in the OGLE-III LMC dataset \citep{Udalski2008} is two orders of magnitude larger than previous surveys.

 \item {\it Physical Model Fits.}  For detached eclipsing binaries with MS primaries and known distances, we can utilize stellar evolutionary tracks to estimate the ages $\tau$ and component masses $M_1$ and $M_2$ based solely on the observed photometric light curves (\S3.1\,-\,3.2).  For the 18 definitive MS\,+\,pre-MS eclipsing binaries, we measured primary masses $M_1$~=~6\,-\,16\,\Msun, secondary masses $M_2$~=~0.8\,-\,2.4\,\Msun\ ($q$~=~0.07\,-\,0.36), and ages $\tau$~=~0.6\,-\,8 Myr (\S3.3).  We investigated multiple sources of systemic uncertainties and performed various consistency checks (\S3.4).  Our conclusions that the majority of our reflecting eclipsing binaries have pre-MS companions with extreme mass ratios $q$~$<$~0.25 and young ages $\tau$~$<$~8~Myr are robust.

 \item {\it Association with H\,\textsc{ii} Regions.}  Relative to our total sample of 2,206 B-type MS eclipsing binaries, the coordinates of our 22 reflecting eclipsing binaries are correlated with the positions of star-forming H\,\textsc{ii} regions at the 4.1$\sigma$ significance level (\S4). In addition, our youngest eclipsing binaries with deeper eclipses and larger reflection effect amplitudes are more likely to be associated with bright and/or compact H\,\textsc{ii} regions. These statistics and correlations: (1)~reinforce our conclusions that our reflecting eclipsing binaries contain young, low-mass, pre-MS companions, (2)~demonstrate the reliability of our eclipsing binary models, and (3)~provide powerful diagnostics for the expansion velocities $\langle v \rangle _{\rm H\,II}$~$\approx$~10\,-\,30~km~s$^{-1}$ and long-term dynamical evolution of H\,\textsc{ii} regions.

 \item {\it Intrinsic Close Binary Statistics.}  We performed detailed Monte Carlo simulations to generate synthetic light curves for a large population of eclipsing binaries (\S5.1\,-\,5.3).  Only ${\cal P}_{\rm refl}$~$\approx$~0.7\% of B-type MS stars with low-mass companions have the necessary ages and orientations to produce detectable eclipses and reflection effects (\S5.4).  Hence, $F$~=~(2.0\,$\pm$\,0.6)\% of B-type MS stars have companions with extreme mass ratios $q$~=~0.06\,-\,0.25 and short orbital periods $P$~=~3.0\,-\,8.5~days.  This is $\approx$10 times larger than that observed around solar-type MS stars in the same period and mass-ratio interval (\S6.1).  Our analysis represents the first direct measurement for the fraction of B-type MS stars with close, low-mass, non-degenerate stellar companions.

 \item {\it Implications for Binary Formation.}  The lack of close extreme mass-ratio companions to solar-type MS stars, commonly known as the brown dwarf desert, is probably because such companions migrated inward at the time of formation in the circumbinary disk and merged with the primary \citep{Armitage2002}. Because massive binaries have rapid disk photoevaporation timescales and larger orbital angular momenta, our extreme mass-ratio eclipsing binaries could therefore stabilize into short orbits without merging (\S6.2). Close binaries with extreme mass ratios may have formed either through: (1)~less efficient competitive accretion in the circumbinary disk \citep{Bate1997}, (2)~tidal capture while the secondary is still a large, convective pre-MS star \citep{Press1977,Moeckel2007}, and/or (3)~Kozai cycles with a tertiary component and subsequent tidal friction between the inner B-type MS + low-mass pre-MS inner binary \citep{Kiseleva1998}.   

\item {\it Implications for Binary Evolution.}  B-type MS stars with closely orbiting low-mass companions $q$~=~0.1\,-\,0.3 can evolve to produce Type~Ia supernovae, low-mass X-ray binaries, and millisecond pulsars \citep{Kiel2006,Ruiter2011}.  We find more close, low-mass companions to B-type MS stars than is typically assumed in binary population synthesis (\S6.3).   If this result holds at slightly longer orbital periods, we anticipate more progenitors of Type Ia supernovae and low-mass X-ray binaries than originally predicted. 

\end{enumerate}

This research was funded by the National Science Foundation under grant AST-1211843.  We acknowledge use of publicly available data from the Optical Gravitational Lensing Experiment \citep{Udalski2008,Graczyk2011} and the Magellanic Cloud Emission Line Survey \citep{Smith2005}.  We thank the referee, A. Pr{\v s}a, for helpful comments and suggestions that improved the quality of the manuscript.  We alo thank N.~Evans, J.~Najita, S.~Naoz, C.~Badenes, I.~Czekala, C.~Faesi, and D.~Nelson for useful discussions.

\bibliographystyle{apj}                       
\bibliography{bibliography}

\end{document}